\documentclass[fleqn,usenatbib]{mnras}
\usepackage[english]{babel}
\usepackage[utf8]{inputenc}

\DeclareRobustCommand{\VAN}[3]{#2}
\let\VANthebibliography\thebibliography
\def\thebibliography{\DeclareRobustCommand{\VAN}[3]{##3}\VANthebibliography}

\usepackage{amssymb}
\usepackage{amsmath}
\usepackage{graphicx}
\usepackage{subfig}
\usepackage{color}
\usepackage{float}
\usepackage{multirow}
\usepackage{newtxtext,newtxmath}

\definecolor{gaolvDarkGray}{gray}{0.5}

\newcommand{\skeleton}[1]{%
\ifx\showskeleton\undefined
\else
\vspace{1em}\noindent$\bigstar${\color{gaolvDarkGray}\emph{#1}}\newline%
\fi
}
\newlength{\subfiglen}

\newcommand{\unit}[1]{\,\mathrm{#1}}

\newcommand{\orcid}[1]{\href{https://orcid.org/#1}{\includegraphics[width=10pt]{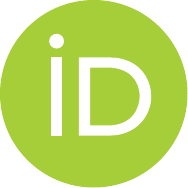}}}

\title[Effects of $q$ on ray-tracing for neutron stars]{Effects of the treatment of the mass quadrupole moment on ray-tracing applications for rapidly-rotating neutron stars}
\author[G. A. Oliva and F. Frutos--Alfaro]{%
G. André Oliva$^{\orcid{0000-0003-0124-1861} 1,2}$\thanks{E-mail: \href{mailto:andree.oliva@uni-tuebingen.de}{andree.oliva@uni-tuebingen.de} (GAO);\newline \href{mailto:francisco.frutos@ucr.ac.cr}{francisco.frutos@ucr.ac.cr} (FF-A)} and
Francisco Frutos--Alfaro,$^{\orcid{0000-0001-8875-5480} 2}$\footnotemark[1]
\\
$^{1}$Institute for Astronomy and Astrophysics, University of Tübingen, Auf der Morgenstelle 10, D-72076 Tübingen, Germany\\
$^{2}$Space Research Center (CINESPA), School of Physics, University of Costa Rica, 11501 San José, Costa Rica.
}

\date{Accepted XXX. Received YYY; in original form ZZZ}

\pubyear{2020}

\makeatletter

\let\jnl@style=\rm
\def\ref@jnl#1{{\jnl@style#1}}

\def\aj{\ref@jnl{AJ}}                   
\def\actaa{\ref@jnl{Acta Astron.}}      
\def\araa{\ref@jnl{ARA\&A}}             
\def\apj{\ref@jnl{ApJ}}                 
\def\apjl{\ref@jnl{ApJ}}                
\def\apjs{\ref@jnl{ApJS}}               
\def\ao{\ref@jnl{Appl.~Opt.}}           
\def\apss{\ref@jnl{Ap\&SS}}             
\def\aap{\ref@jnl{A\&A}}                
\def\aapr{\ref@jnl{A\&A~Rev.}}          
\def\aaps{\ref@jnl{A\&AS}}              
\def\azh{\ref@jnl{AZh}}                 
\def\baas{\ref@jnl{BAAS}}               
\def\bac{\ref@jnl{Bull. astr. Inst. Czechosl.}}
\def\caa{\ref@jnl{Chinese Astron. Astrophys.}}
\def\cjaa{\ref@jnl{Chinese J. Astron. Astrophys.}}
\def\icarus{\ref@jnl{Icarus}}           
\def\jcap{\ref@jnl{J. Cosmology Astropart. Phys.}}
\def\jrasc{\ref@jnl{JRASC}}             
\def\memras{\ref@jnl{MmRAS}}            
\def\mnras{\ref@jnl{MNRAS}}             
\def\na{\ref@jnl{New A}}                
\def\nar{\ref@jnl{New A Rev.}}          
\def\pra{\ref@jnl{Phys.~Rev.~A}}        
\def\prb{\ref@jnl{Phys.~Rev.~B}}        
\def\prc{\ref@jnl{Phys.~Rev.~C}}        
\def\prd{\ref@jnl{Phys.~Rev.~D}}        
\def\pre{\ref@jnl{Phys.~Rev.~E}}        
\def\prl{\ref@jnl{Phys.~Rev.~Lett.}}    
\def\pasa{\ref@jnl{PASA}}               
\def\pasp{\ref@jnl{PASP}}               
\def\pasj{\ref@jnl{PASJ}}               
\def\rmxaa{\ref@jnl{Rev. Mexicana Astron. Astrofis.}}%
\def\qjras{\ref@jnl{QJRAS}}             
\def\skytel{\ref@jnl{S\&T}}             
\def\solphys{\ref@jnl{Sol.~Phys.}}      
\def\sovast{\ref@jnl{Soviet~Ast.}}      
\def\ssr{\ref@jnl{Space~Sci.~Rev.}}     
\def\zap{\ref@jnl{ZAp}}                 
\def\nat{\ref@jnl{Nature}}              
\def\iaucirc{\ref@jnl{IAU~Circ.}}       
\def\aplett{\ref@jnl{Astrophys.~Lett.}} 
\def\apspr{\ref@jnl{Astrophys.~Space~Phys.~Res.}}
\def\bain{\ref@jnl{Bull.~Astron.~Inst.~Netherlands}}
\def\fcp{\ref@jnl{Fund.~Cosmic~Phys.}}  
\def\gca{\ref@jnl{Geochim.~Cosmochim.~Acta}}   
\def\grl{\ref@jnl{Geophys.~Res.~Lett.}} 
\def\jcp{\ref@jnl{J.~Chem.~Phys.}}      
\def\jgr{\ref@jnl{J.~Geophys.~Res.}}    
\def\jqsrt{\ref@jnl{J.~Quant.~Spec.~Radiat.~Transf.}}
\def\memsai{\ref@jnl{Mem.~Soc.~Astron.~Italiana}}
\def\nphysa{\ref@jnl{Nucl.~Phys.~A}}   
\def\physrep{\ref@jnl{Phys.~Rep.}}   
\def\physscr{\ref@jnl{Phys.~Scr}}   
\def\planss{\ref@jnl{Planet.~Space~Sci.}}   
\def\procspie{\ref@jnl{Proc.~SPIE}}   

\makeatother

\begin{document}
\label{firstpage}
\pagerange{\pageref{firstpage}--\pageref{lastpage}}
\maketitle

\begin{abstract}
	The Neutron Star Interior Composition Explorer (NICER) mission has provided a unique opportunity to constrain the equation of state of neutron stars by using the technique of pulse-profile modelling. This technique requires accurate and efficient ray tracing, that in turn requires a robust representation of the spacetime around a neutron star. Several exact and approximate metrics have been proposed, and used, to perform ray tracing around neutron stars, with both moderate and fast rotation. In this paper, we perform a comparison between several of these metrics, when used for ray tracing. We calculate the shape of the neutron star as seen by a distant observer using two different surface formulae, the thermal spectrum and pulse profiles from circular and crescent-shaped hotspots, for four configurations of pulsars with rotation rates ranging from 622 to 1000 Hz, and using both a moderate and a stiff equation of state to include realistic and extreme cases. We find small differences between the metrics for rotation frequencies starting at $\sim 700\unit{Hz}$ that could theoretically be used for constraining the quadrupole moment or the spacetime models. We also determine the practicality of use of each metric in larger-scale applications such as pulse-profile modelling.

\end{abstract}


\section{Introduction}
\skeleton{Observational motivation}
The problem of finding a solution of the Einstein field equations for a neutron star is still unsolved, in part because of the lack of sufficient knowledge about the equation of state of dense nuclear matter. With the advent of the Neutron Star Interior Composition Explorer (NICER) mission, it is now possible to perform high precision measurements of the mass and radius of a neutron star, which also constraint its equation of state \citep{2019ApJ...887L..22R, Miller_2019}.

Hot emitting regions on the surface of a neutron star, or hotspots, are responsible for the periodic signal observed from pulsars. Pulse-profile modelling is a technique that uses the general relativistic effect on the X-ray thermal emission from hotspots in order to measure neutron star parameters \citep[see, e.g.,][]{2016RvMP...88b1001W, Bogdanov_2019}. Isolated pulsars of rotation frequencies $\sim 200\unit{Hz}$ were selected by the NICER group in order to constrain the mass-radius relation of neutron stars with this technique, by fitting a large number of light-curve models produced with different parameters to the observational data \citep{2019ApJ...887L..21R, Miller_2019}. Rapidly rotating pulsars, however, are known to exist (PSR J1748-2446ad has a rotation frequency of 716\,Hz), and their rotation causes deformation and frame dragging that in turn affect their emission, and these effects should be taken into account into future observational studies of such targets.


\skeleton{State of the art: ray tracing, hotspots and thermal spectrum}
\cite{1983ApJ...274..846P} used the Schwarzschild metric for modelling the light curve of the radiation emitted by a hotspot. In \cite{1998ApJ...499L..37M}, \cite{2003MNRAS.343.1301P} and \cite{2006MNRAS.373..836P}, the approximate effects of the Doppler effect by rotation were added to the Schwarzschild metric to the pulse profiles. \cite{2007ApJ...654..458C} tried a completely numerical approach, by using a numerical metric to integrate the geodesic equations and calculate the light curves. \cite{2007ApJ...663.1244M}, in the other hand, improved the Schwarzschild+Doppler approximation by adding an approximation to the oblateness of the neutron star.

\cite{2012ApJ...753..175B} used an approximate metric \citep[developed by][hereafter GB]{2006CQGra..23.4167G} that included quadrupole moment in order to study how it affects the apparent shape of a rapidly rotating neutron star as seen by a distant observer. They used values of the quadrupole moment fitted empirically by \cite{1999ApJ...512..282L} for several equations of state. Using the same metric, \cite{2014ApJ...792...87P} calculated the pulse profiles, and did extensive comparisons with the results obtained with the Schwarzschild+Doppler approximation and the oblate Schwarzschild+Doppler approximation. A similar comparison (among many other ray-tracing applications) was performed in \cite{2018ApJ...863....8P}, but using the metric of \cite{2014ApJ...791...78A}, which is a fitted expansion of the metric of \cite{1976ApJ...204..200B}. \cite{2015ApJ...799...22B} used the GB metric in order to calculate corrections to the thermal spectrum due to rotation. In \cite{2018A&A...615A..50N}, a theoretical framework for ray tracing for rapidly-rotating neutron stars was introduced.  The effects of the atmosphere were incorporated to a numerical metric approach by \cite{2018ApJ...855..116V}.

Some of the studies mentioned above contain the classical model of two antipodal hotspots. In \cite{2019ApJ...887L..21R} and \cite{Miller_2019}, however, more complex configurations were explored, like annuli, crescent-shaped and oval spots, with the result that the antipodal configuration was strongly disfavoured in the cases analysed.

\skeleton{State of the art: metrics for the exterior spacetime of a NS}
Apart from the approximate spacetime models mentioned already, there are additional exact and approximate metrics that have been proposed as a description of the exterior spacetime of a spinning neutron star \citep[among others,][]{QM, Manko_1992,  2006CQGra..23.4167G, PhysRevD.86.064043, 2016IJAA....6..334F, 10.1093/mnras/stx019, Frutos2019}. Although such metrics converge for slow rotation and small deformation, an assessment and comparison of their practical use in ray-tracing applications for rapidly rotating neutron stars has not been performed until now. The expected small differences between them may give rise to small uncertainties when used in pulse profile modelling and fitting of the spectrum.

\skeleton{Aims of the paper, structure}
The present study aims to estimate these small differences for rapidly rotating neutron stars, and establish an optimal metric that minimizes computing resources while keeping an adequate accuracy. We also expand the parameter range presented in previous similar studies and explore the coupled effects of rotation and deformation on ray-tracing applications. The present paper is structured as follows. Sect. \ref{s: metrics} gives a general introduction to the metrics used, Sect. \ref{S: nsparam} deals with the chosen physical objects for this study and their parameters. A direct comparison between a numerical solution of the Einstein field equations and the analytical metrics is presented in Sect. \ref{s: rns-vs-analytical}. In sections \ref{S: raytracing}--\ref{S: ppm}, we present several ray-tracing applications: light scattering, determination of the shape of the neutron star, the thermal spectrum and pulse profiles. In each section, the methods and results are detailed. Section \ref{s: summary} offers a summary and general conclusions. Unless stated otherwise, we use geometrized units ($G=c=1$) throughout the paper.

\section{Exterior Kerr-like metrics}\label{s: metrics}

In the case of analytical solutions of the Einstein field equations, the problem is typically divided into an interior and an exterior solution. There are some exact exterior solutions, and some approximate analytical solutions, a subset of which we review in the following sections, and use throughout this study.

\skeleton{Definition of the quadrupole moment}
In Newtonian gravity, an expansion up to order $r^{-3}$ of the gravitational potential outside a static spheroid of equatorial radius $R$ and mass $M$ is given by \citep[see, e.g., ][ sect. 3.3.4]{capderou2014}:
\begin{equation} \label{eq: Phi newt}
\Phi = -\frac{M}{r} + \frac{q}{r^3} P_2(\cos\theta),
\end{equation}
where $q$ is the mass quadrupole (in this case, the difference between the moments of inertia along the polar and equatorial axes; with dimensions of $MR^2$), and $P_2(\cos\theta) = (3\cos^2 \theta - 1)/2$ is the Legendre polynomial of second order.

In the general relativistic case, consider a neutron star modelled as a rotating spheroid of mass $M$, angular momentum $J \equiv Ma$, and mass quadrupole moment ($q>0$ for an oblate object). In principle, the mass quadrupole could be determined from the mass, angular momentum and equation of state that describes the interior of the neutron star. The most general solution of the Einstein field equations for the exterior (vacuum) spacetime around such an object has the general form
\begin{multline} \label{eq: gensol}
ds^2 = -V(r,\theta)\, dt^2 + 2 W(r,\theta)\, dt \, d\phi + X(r,\theta)\, dr^2\\
 + Y(r,\theta) \,d\theta^2 + Z(r,\theta)\, d\phi^2,
\end{multline}
where the potentials $V, W, X, Y$ and $Z$ depend on $M$, $a$ and $q$ as well. In the limit $r\to \infty$, the potential $V$ might be expanded as \citep[see e.g.][]{Frutos2019}:
\begin{equation}
	V \approx 1 - 2\Phi - 2\frac{Ma^2}{r^3} \cos^2\theta,
\end{equation}
where the last term is due to rotation (matching with the Kerr solution when $q=0$). The second order moment (coefficient of the $r^{-3}$ terms) is then
\begin{equation} \label{eq:M2 def}
	M_2 = -q - Ma^2.
\end{equation}

\skeleton{Exact metrics}
There are some exact solutions that contain the required parameters. In this study, we focus in particular on two exact metrics: the Quevedo--Mashhoon (QM) metric \citep{QM} and the Manko--Novikov (MN) metric \citep{Manko_1992}. The QM metric is a generalization of the Kerr and Erez--Rosen spacetimes, i.e., it reduces to these metrics when the mass quadrupole moment and rotation vanish, respectively. The MN metric is also a generalization of the Kerr metric, and it contains arbitrary mass multipole moments. Both metrics have a complex analytic representation (see Appendix \ref{s: appendix-metrics}), and their practical use in applications of ray-tracing is explored in Sect. \ref{S: raytracing}.

\skeleton{Approximate metrics}
There are several approximate metrics that have a simpler analytical form. \cite{1968ApJ...153..807H} (hereafter HT) solved approximately the Einstein field equations for terms up to order $J^2$ and $q$. The Glampedakis--Babak (GB) metric \citep{2006CQGra..23.4167G}, used in previous studies of ray-tracing applications, is a perturbed version of the Kerr metric using the corrected form of Abramowicz for the Hartle--Thorne metric. The approximate metric presented in \cite{2016IJAA....6..334F} (hereafter, Fru16) also contains the Kerr(-Newman) metric, and an approximation to the Erez--Rosen metric up to second order in quadrupole moment. A transformation to the expanded Hartle--Thorne metric was found, and more multipole moments can be added to the approximation \citep{Frutos2019}.

\skeleton{Redirection to appendix}
The metric potentials $V$, $W$, $X$, $Y$ and $Z$ are given for each of these metrics in Appendix \ref{s: appendix-metrics}, as well as a multipole analysis (given that not all the metrics use the same definitions of the quadrupole moment) and the necessary transformations to the definition given by equation \eqref{eq:M2 def}.

\section{Choice of neutron star parameters} \label{S: nsparam}
\begin{table*}
\begin{tabular}{| c || c | c || c | c || c | c || c | c | c | c |}
\hline
\multirow{2}{*}{config.}& \multicolumn{2}{c || }{Mass} & \multicolumn{2}{c ||}{Rotation} & \multicolumn{2}{c ||}{Deformation} & \multicolumn{4}{c |}{Model parameters} \\ \cline{2-11}
 & $\mathrm{[M_\odot]}$ & $M/R$ & freq. $\mathrm{[Hz]}$ & $a/R$ & $R_p/R$ & $q/R^3$ & $R\unit{[km]}$ &  $\rho_c \unit{[g/cm^3]}$ & EoS & condition \\ \hline\hline
{\tt BWFX} & 1.82 & $0.2836$ & 622 & 0.05424 & 0.9675 & 0.001450 & 9.487 & $3.2 \cdot 10^{15}$ & FPS & max. mass \\ \hline
{\tt SHFT} & 1.44 & 0.1854 & 716 & 0.06129 & 0.8972 & 0.002829 & 11.43 & $1.25 \cdot 10^{15}$ & FPS & typ. mass \\ \hline
{\tt KAFT} & 1.41 & 0.1648 & 1000 & 0.08664 & 0.7568 & 0.005277 & 12.65 & $1.1\cdot 10^{15}$ & FPS & typ. mass \\ \hline
{\tt KALN} & 2.72 & 0.2003 & 1000 & 0.1429 & 0.5703 & 0.010016 & 20.05 & $6.1\cdot 10^{14}$ & L & min. mass \\ \hline
\end{tabular}

\caption{Parameters of the pulsars considered in this study. From left to right, the columns indicate: the name of the configuration, the (gravitational) mass of the neutron star in solar masses, the mass expressed in geometrized units of the equatorial radius $R$, the frequency of rotation, the specific angular momentum in units geometrized units of $R$, the ratio of the polar to equatorial radii, the mass quadrupole moment in geometrized units of the equatorial radius, the equatorial radius, the central density, the name of the equation of state used (following the convention in {\tt RNS} and the condition on the mass that defines the model.}
\label{t: config}
\end{table*}

\skeleton{Method}
We are interested in millisecond pulsars, since high rotation rates can produce high deformation, and therefore, higher quadrupole moment and a better opportunity to find disagreement between the results yielded by the use of different metrics. We used the program Rapidly Rotating Neutron Star ({\tt RNS}) \citep[][]{1995ApJ...444..306S} in order to model the parameters of several rapidly rotating neutron stars. {\tt RNS} solves the Einstein field equations numerically by first computing a non-rotating equilibrium configuration (by solving the Tolman–Oppenheimer–Volkoff equation) and iteratively adding rotation with a Green function approach \citep[for more details, see][]{1989MNRAS.237..355K, 1992ApJ...398..203C}.

\skeleton{Selection of astrophysical objects}
We selected the following configurations: {\tt BWFX} corresponds to the Black Widow pulsar (PSR B1957+20); {\tt SHFT} corresponds to PSR J1748-2446ad, the fastest spinning pulsar known to date (rotation frequency: $716\unit{Hz}$), and {\tt KAFT} and {\tt KALN}, that correspond to hypothetical pulsars that rotate with a frequency of $1000\unit{Hz}$. The full parameters corresponding to those configurations are shown in Table \ref{t: config}.

\skeleton{Equations of state}
Two classic equations of state were used: FPS \citep{1993PhRvL..70..379L} and L \citep[described in ][]{1977ApJS...33..415A}. The equation of state L is based on the mean-field approximation and is unrealistically stiff; it is used here only to consider an extreme case of deformation with a high rotation frequency. The equation of state FPS is more realistic because it produces more moderately stiff configurations, and it is therefore used for the other cases considered here.

The Black Widow pulsar has a rotation frequency of $622\unit{Hz}$, and its mass has been constrained to $>1.66\unit{M_\odot}$ \citep{2011ApJ...728...95V}. For this reason, we chose the configuration that yielded the most massive neutron star using the equation of state FPS. The configurations {\tt SHFT} and {\tt KAFT} consider a typical neutron star mass of $1.4\unit{M_\odot}$, and the extreme case of {\tt KALN} was chosen with the minimum mass possible, which produced the maximum deformation. The equation of state L yields for the configuration {\tt KALN} a value of the mass that is probably unrealistically high; however, we keep the configuration as an extreme case as a reference only.

\section{Direct numerical comparison of the metric components} \label{s: rns-vs-analytical}
\begin{figure*}
    \centering
    \includegraphics[width=\textwidth]{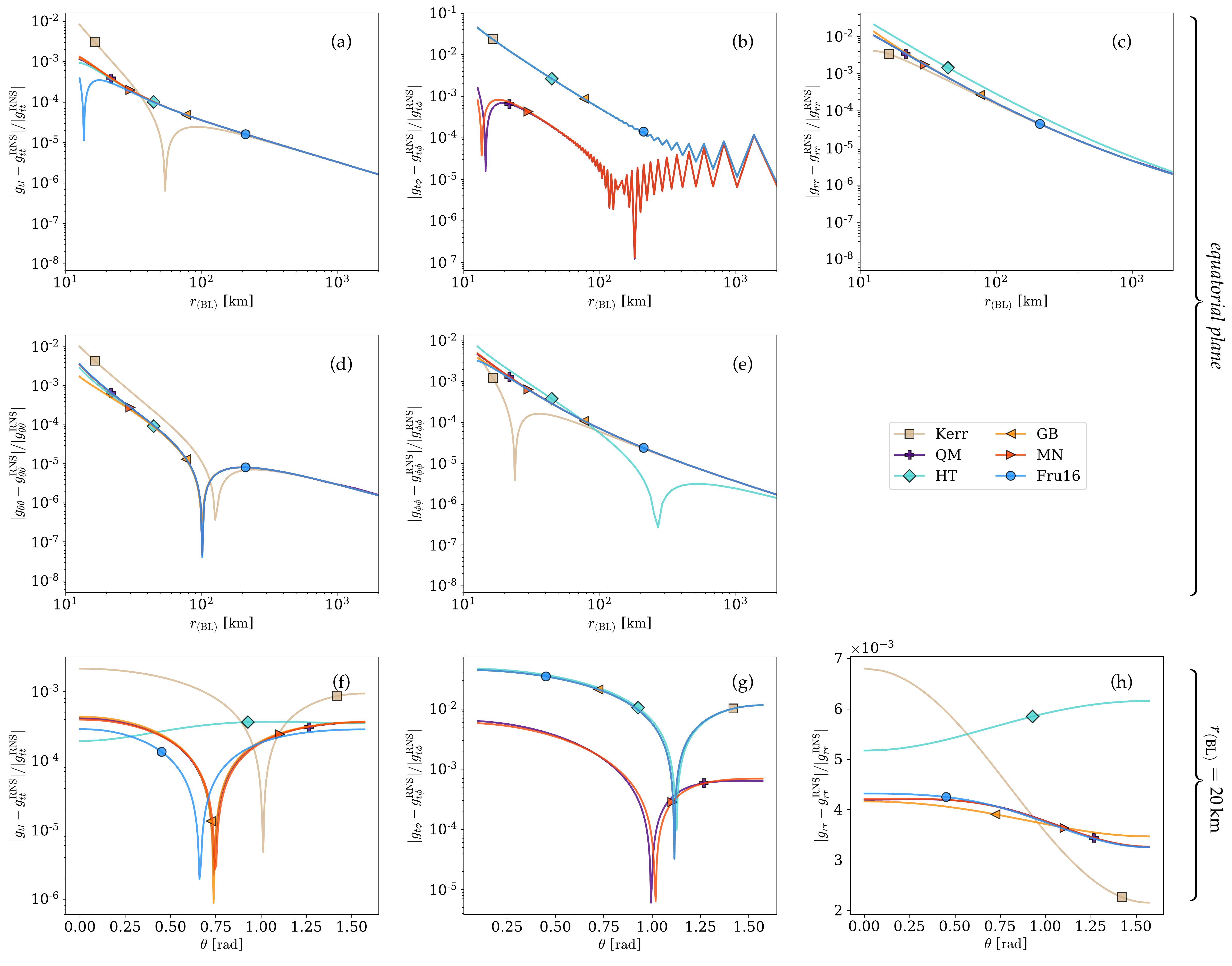}

	\caption{Direct numerical comparison of the metric components against the exterior numerical solution provided by {\tt RNS}. Panels \emph{a}--\emph{e} show the radial behavior and correspond to the equatorial plane ($\theta=0$), while panels \emph{f}--\emph{h} show the angular behavior and are evaluated at $r=20\unit{km}$. All radii are expressed in Boyer--Lindquist (BL) coordinates. The configuration {\tt KAFT} was used for this numerical comparison.}
	\label{f: comp_rns}
\end{figure*}

\skeleton{Setup}
We present here a comparison between the components of the exterior numerical metric produced by {\tt RNS} with its analytical counterparts described in Sect. \ref{s: metrics}. The configuration {\tt KAFT} was given as an input to {\tt RNS}, as well as a grid of 401 cells in the radial direction and 201 cells in the polar direction.

\subsection{Obtaining parameters and coordinates from the numerical solution}
\skeleton{r}
The Komatsu-Eriguchi-Hachisu scheme used by {\tt RNS} uses the general form of a stationary and axisymmetric metric in quasi-isotropic coordinates. In order to compare the numerical solution with analytical metrics written in Boyer-Lindquist coordinates, we used the transformation provided by \cite{PhysRevD.54.1403}:
\begin{equation}
r_\text{BL} = r_\text{I} \left( 1 + \frac{M_\text{[g]} + a_\text{[g]} }{2r_\text{I}} \right) \left( 1 + \frac{M_\text{[g]} - a_\text{[g]}}{2r_\text{I}} \right) ,
\end{equation}
where $r_\text{I}$ is the radial coordinate in quasi-isotropic coordinates, $r_\text{BL}$ is the radial coordinate in Boyer-Lindquist coordinates, and the subindex $\text{[g]}$ emphasizes the use of geometrized units.

\skeleton{a}
Given the angular frequency of rotation $\Omega$, the mass $M$, and the moment of inertia $\mathcal I$ computed by {\tt RNS}, all in CGS units, the specific angular momentum $a$ in geometrized units of $\mathrm{cm}$ was computed as
\begin{equation}
a_\text{[g]} = \left(\frac{\mathcal I \Omega}{M c} \right)_\text{[cgs]}.
\end{equation}

\skeleton{q}
For the mass quadrupole moment, we followed the discussion in, e.g., \cite{2012PhRvL.108w1104P} and \cite{2014ApJ...781L...6D}: {\tt RNS} computes the second order multipole, and by an asymptotic expansion of the potentials that form the numerical solution, it is possible to isolate the contribution to the $r^{-3}$ terms that corresponds to the mass quadrupole moment, by subtracting the contribution of rotation, in a way that is analogous to equation \eqref{eq:M2 def}.

\subsection{Results of the comparison}
\skeleton{Description of the plots}
Figure \ref{f: comp_rns} shows the fractional difference between each one of the metrics considered in this study (see Sect. \ref{s: metrics}) and the numerical solution by {\tt RNS}, for the nonzero components of the metric evaluated at the equatorial plane (panels \emph{a}--\emph{e}), and the angular behavior of some components evaluated at a fixed distance of $r=20\unit{km}$ (panels \emph{f}--\emph{h}). As a reference, the Kerr solution is also shown.

In all cases, the solutions tend to agree more with increasing radial position. The asymptotic features in all the curves correspond to a change of sign in the fractional difference. The component $g_{t\phi}$ (panels \emph{b} and \emph{g}) show a clear distinction between the exact metrics with mass quadrupole moment (QM and MN) and the approximations. In the $g_{tt}$ component (panels \emph{a}, \emph{f}), QM, MN, HT and GB are very similar. The Fru16 metric becomes marginally better close to the surface of the neutron star, while the Kerr metric becomes insufficient in the same region. In the $g_{rr}$ component, HT is marginally worse than the rest of the metrics. We expect the differences in those three components to dominate over the other spacial components. These results are consistent with what \cite{2004MNRAS.350.1416B} found when comparing the components of the metric by \cite{PhysRevD.61.081501} and the Kerr metric with the numerical metric by {\tt RNS}, as well as the findings by \cite{10.1093/mnras/stx019}, who compared the numerical solution with another analytical solution and the HT metric.

\skeleton{Conclusion: metrics are only `consistent'}
Considering the information in the plots, and the expected numerical accuracy of the solution provided by {\tt RNS}, we conclude that all the analytical metrics in Sect. \ref{s: metrics} are consistent with the numerical solution. However, with this comparison, we cannot establish an order of accuracy between the metrics at large distances, or an order of accuracy at smaller distances other than the mentioned general distinction between exact and approximate metrics.

\section{Ray tracing for light scattering} \label{S: raytracing}

\subsection{Method}
\skeleton{Ujti}
For the ray-tracing applications, we developed a software package called {\tt Ujti}\footnote{`path' in Nawat language. The source code is available from {\tt http://cinespa.ucr.ac.cr}}, that first calculates symbolically the geodesic equations given a axially-symmetric metric, and then solves them numerically. We used an early version of the program in \cite{oliva15}. We start by calculating a general expression of the geodesic equations for the metric \eqref{eq: gensol} in Boyer-Lindquist coordinates, yielding expressions that are dependent on the potentials $V,W,X,Y,Z$ and their derivatives. These expressions are then calculated for each metric, and numerically evaluated when solving the geodesic equations.

\skeleton{Geodesic equations}
Using the Runge-Kutta-Fehlberg method (4th order, with an error estimator of the 5th order) as described in \cite{burden2011numerical}, we compute the numerical solution of the geodesic equations

\begin{equation}
	\frac{d^2 x^\kappa}{d\ell^2} = -\Gamma^{\kappa}_{\mu \nu} \frac{dx^\mu}{d\ell}\frac{dx^\nu}{d\ell}\text{.}
\end{equation}

\skeleton{Initial conditions}
We set the wave vector $k^\mu$ of a photon that follows a null geodesic to $dx^\mu/d\ell$. The affine parameter $\ell$ can be fixed by considering the component $k^0 = \omega$ and then integrating to set $\ell \equiv t/\omega$, i.e., the affine parameter is taken as being proportional to the proper wavelength of the photon.

In all the applications considered in this paper, the geodesics are released parallel to each other from  distant points. The null geodesic condition, $k^\mu k_\mu = 0$, is also imposed to the initial wave vector by normalizing it.

\subsection{Light scattering results and discussion} \label{s: raytr-results}

\begin{figure}
	\centering
	\includegraphics[width=\columnwidth]{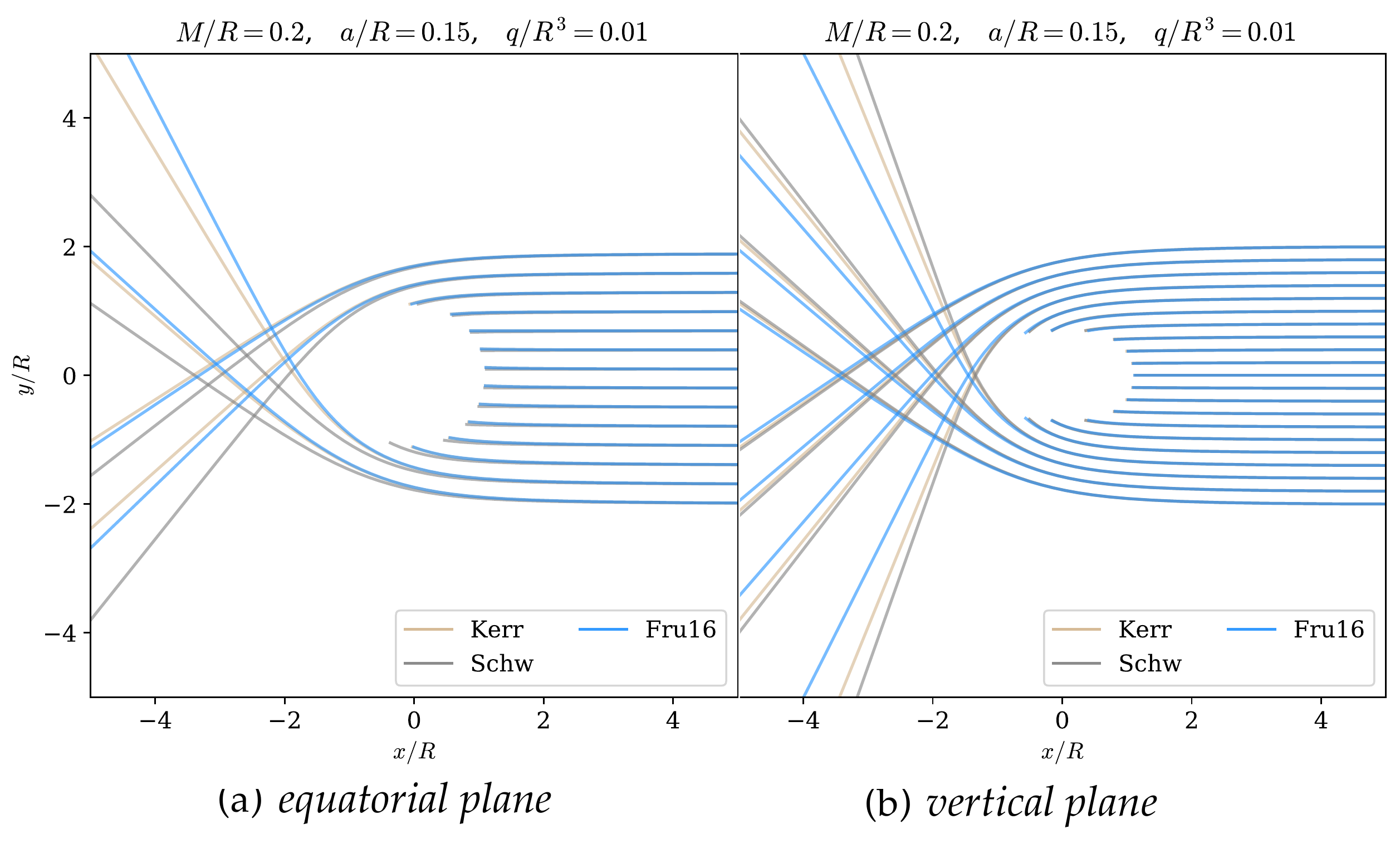}
	\caption{Effect of the quadrupole moment as defined in the Fru16 metric.}
	\label{f: raytr-aq}
\end{figure}

\begin{figure*}
	\includegraphics[width=\textwidth]{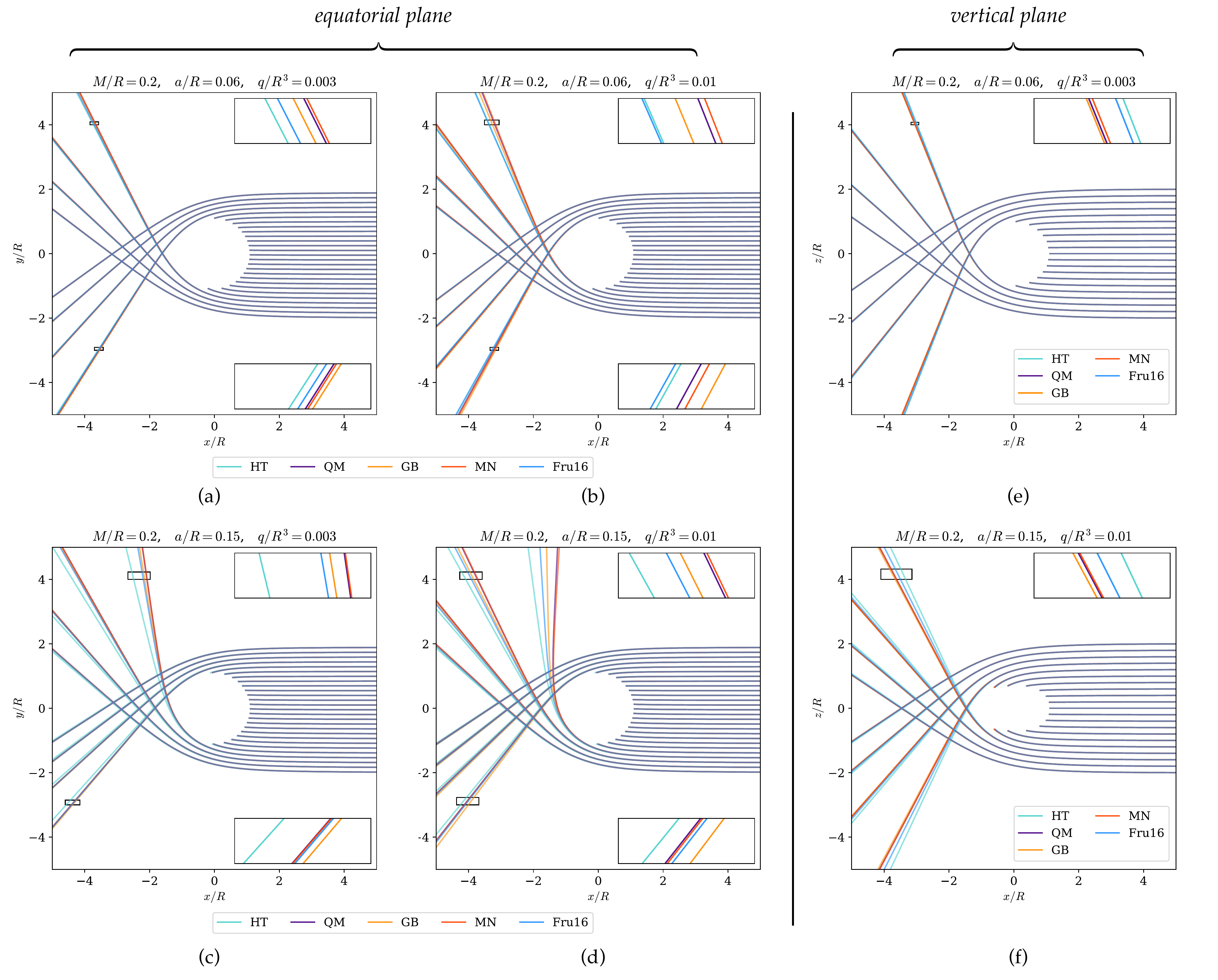}
	\caption{Light scattering for different metrics. The vertical plane considered in (e) and (f) is $y/R = 0.2$. Inset in each graph is a zoomed-in portion of the corresponding rectangle.}
	\label{f: raytr}
\end{figure*}

\subsubsection{The relative importance of $a$ and $q$}
\skeleton{Importance of a and q: intro}
In order to explore the relative importance of $a$ and $q$ in light scattering, we trace rays with the Fru16 metric (that contains both parameters), the Kerr metric (which is the limit when $q=0$), and the Schwarzschild metric (which is the limit when both $a$ and $q$ vanish). We took values of $M/R$, $a/R$ and $q/R^3$ that are consistent with the extreme values in Table \ref{t: config}, namely $M/R=0.2$, $a/R=0.15$ and $q/R^3 = 0.01$. The ratio of polar to equatorial radii $R_p/R$ was taken to be $0.9$ for $q/R^3=0.003$.

\skeleton{Importance of a and q: results}
Figure \ref{f: raytr-aq} shows the results of this exercise. In the equatorial plane, the Schwarzschild metric yields, as expected, symmetric results, while the Kerr metric shows asymmetry product of frame dragging. The Fru16 metric also exhibits the same frame dragging as the Kerr metric, but the mass quadrupole moment increases slightly the deflection angle, so that the geodesics intersect at a slightly closer position from the neutron star.

The vertical plane in the same figure corresponds to the projection of the geodesics onto the $y/R=0.2$ plane; the plane $y=0$ was not used because it contains the coordinate singularity of the $z$ axis in spherical-like coordinates. In the vertical plane, the situation is symmetrical respect to the midplane. The parameter $a$ in the Kerr metric has the effect of decreasing the deflection angle, which is decreased even further by the presence of $q$ in the Frutos metric.

\subsubsection{Differences between metrics in ray-tracing}

\skeleton{Overview: extended parameter scan}
A parameter scan was performed for the rest of the metrics by considering $M/R = 0.2$, $a/R = \{0.06, 0.15\}$ and $q/R^3=\{0.003, 0.01\}$. All four combinations of these parameters are shown in Fig. \ref{f: raytr} for the equatorial plane, and only the more extreme cases in the vertical plane. $R_p/R$ (ratio of polar to equatorial radii) was taken as $0.9$ for $q/R^3=0.003$, and $0.6$ for $q/R^3=0.01$.

\skeleton{Effects of a}
With the increase of $a$ (and fixed $q$), frame dragging is stronger and the geodesics in the equatorial plane tend to bend in the direction of rotation, as expected from the well-known behaviour of the null geodesics of the Kerr metric. Again, the vertical plane projections do not exhibit variation with $a$.

\skeleton{Effects of q, difference between metrics}
With the increase of $q$ (and fixed $a$), the deflection angle tends to increase in the equatorial plane, but this increase is metric-dependent. Fru16 and HT underestimate the effect of $q$ with respect to the exact solutions (MN and QM), while GB underestimates it for the geodesics that bend in favor of rotation but overestimates it in the geodesics that bend against rotation. We relate this result to the methods of approximation between metrics (GB uses a linear deviation from Kerr while Fru16 uses a multiplicative approach). In the vertical direction, GB slightly overestimates the effect of $q$, while Fru16 and HT underestimate it.

In general, the null geodesics of the expanded HT metric agree with the rest of the metrics, with the important exception of the case with high $a$ and low $q$, where the solutions are more similar to the Kerr metric and therefore, the Hartle--Thorne approximation starts not to be sufficient. According to Table \ref{t: config}, however, an increase in $a$ is accompanied with an increase in $q$ for the equations of state considered. The differences between the metrics quickly decay with increasing impact parameter, and, from the plots, they are more notable in the regions closer to the edge of the neutron star.

\skeleton{Conclusions}
From these results, we conclude that the approximate metrics both underestimate and overestimate the results of the exact solutions. In most regions, the geodesics of the exact metrics lie in between the ones of GB and Fru16. Those metrics can then be used to do ray-tracing, and their difference can be used to estimate where the solution by the QM and MN metrics should lie. Based on these results as well, we believe that approximate analytical metrics provide better accuracy than a numerical integration of the geodesic equations that uses the numerical solution.

\subsection{Numerical considerations} \label{s: numerical}
\begin{figure}
\includegraphics[width=\columnwidth]{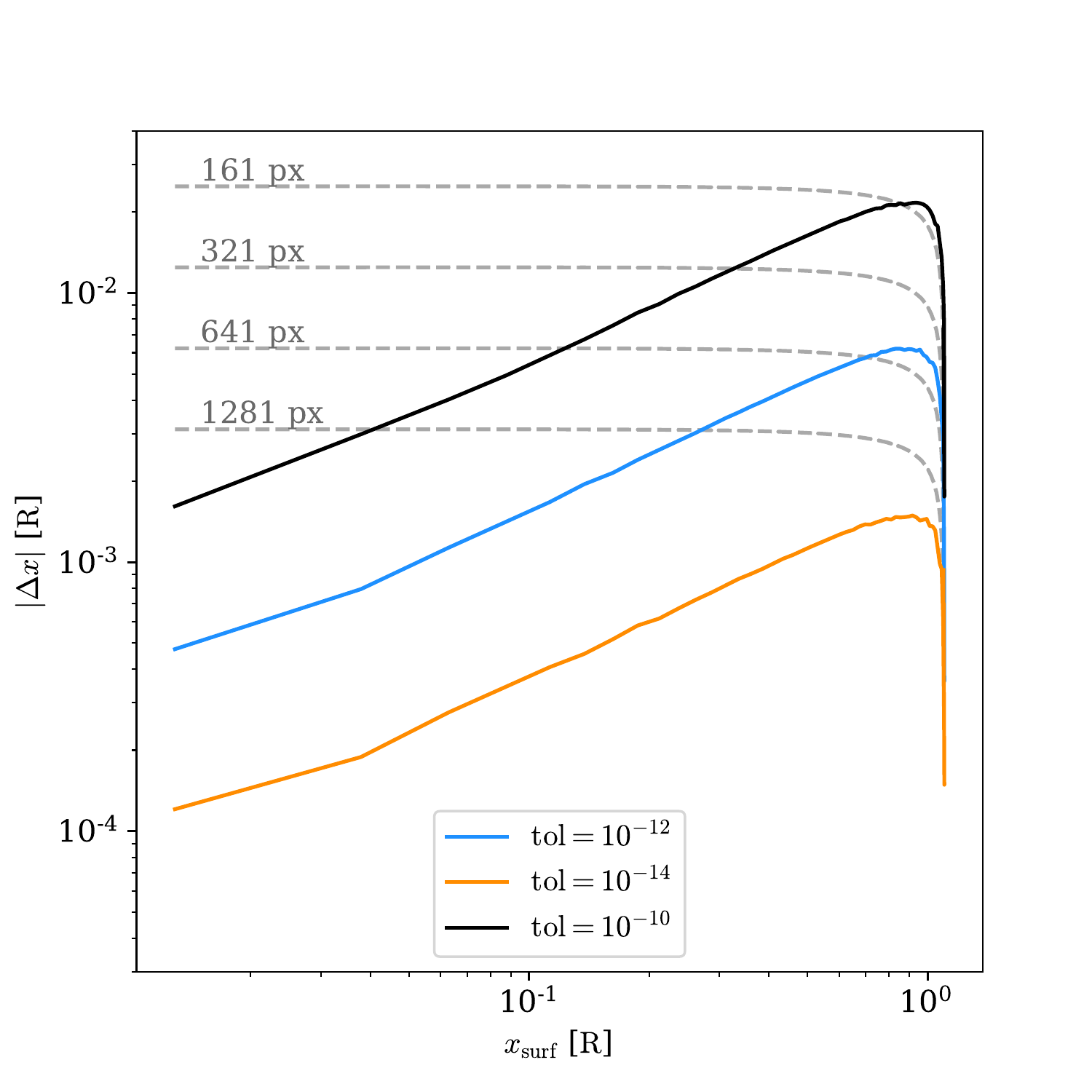}
\caption{Convergence of the light-scattering results for the Fru16 metric. The horizontal axis is the final $x$ position of photons arriving at the surface of the neutron star, and the vertical axis is the difference between this final position and the lowest tolerance (maximum accuracy) run. Plots for the other metrics give a similar result.}
\label{f: convergence}
\end{figure}


\skeleton{Convergence}
Figure \ref{f: convergence} shows a convergence study for the results in the preceding section. Specifically, we focus on the final positions of the photons that land on the surface of the neutron star for the equatorial plane in the case of $M/R = 0.2$, $a/R = 0.06$, $q/R^3 = 0.003$ (Fig. \ref{f: raytr}a), with increasing values of the tolerance in the Runge--Kutta--Fehlberg method, that defines the accuracy of the solution. The horizontal axis corresponds to the final $x$ position, and the vertical axis, to the difference between this final position and the lowest tolerance run ($\mathrm{tol} = 10^{-16}$, i.e., the highest accuracy run). In general, this difference is higher in regions closer to the edge of the neutron star, in which the null geodesics are more affected by the general relativistic effects. The plot also shows the convergence characteristic of the Runge--Kutta--Fehlberg method. Similar plots for other metrics yield the same kind of convergence, and so, differences between metrics are mostly resolution-independent.

The next sections of this paper focus on several applications of ray tracing that require the construction of a grid of initially-parallel geodesics released from a long distance. We are interested in knowing the minimum accuracy (highest tolerance) required so that the initial and final positions of the photon converge within the size of a grid cell (hereafter, ``pixel''). In figure \ref{f: convergence}, dashed lines indicate the distance in $x$ between geodesics that land at the surface of the neutron star, i.e., how the size of one pixel is transformed (lensed) by spacetime for a given grid size.

For the grids that we use in the next sections, a tolerance of $10^{-10}$ is enough to cover the neutron star with $\sim 106$ pixels across (161\,px grid) with at most one pixel of uncertainty;
a tolerance of $10^{-12}$ is enough to cover it with $\sim 212$ pixels across (321\,px grid) with no uncertainty, or $\sim 424$ pixels across (641\,px grid) with one pixel of uncertainty. Finally,  a tolerance of $10^{-14}$ is enough to cover it with $\sim 848$ pixels across (1281\,px grid) with no uncertainty.

Using a very low tolerance not only increases computing time, but also requires too low values of the step size. This causes the numerical integration to stop in the regions close to the $z$ axis when the maximum numerical precision is reached. In the other hand, in determining the shape of hotspots, a too small grid introduces additional uncertainty merely by the discretization of the observer's plane. Considering these elements, we chose to use a tolerance of $10^{-10}$, and a grid of $1281\unit{px}$ across, which means that the uncertainty of the final positions of the photons is of $\pm 4\unit{px}$ across, when comparing our results to observations.

\skeleton{CPU time}
Computing times are critical for ray-tracing applications, a fact that motivates the use of the oblate Schwarzschild+Doppler approximation for pulsars of rotation frequencies of $\sim 200 \mathrm{Hz}$ \citep{Bogdanov_2019}. For the metrics considered in this study, we crudely report the computing times of the (serial) execution of the light scattering applications in this section, in order to assess the practicality of use of each metric in more complex ray-tracing applications. If we call $\tau$ the computing time for tracing a geodesic for the expanded HT metric, the GB metric takes $\approx 1.15 \tau$; the Fru16 metric, $\approx 2.7 \tau$; the MN metric, $\sim 350 \tau$; and the QM metric, $\sim 600 \tau$, meanwhile, the Kerr metric takes $\sim 0.8 \tau$, and the Schwarzschild metric, $\sim 0.3\tau$. The computing times for the exact solutions MN and QM are too high for practical application, which justifies the use of analytical approximations for studies that involve ray-tracing of rapidly-rotating neutron stars. Because of this, in the sections that follow, we restrict ourselves to study the applications of the approximate metrics (expanded HT, GB and Fru16).

\section{Shape of the neutron star} \label{s: shape}

\begin{figure}
\includegraphics[width=\columnwidth]{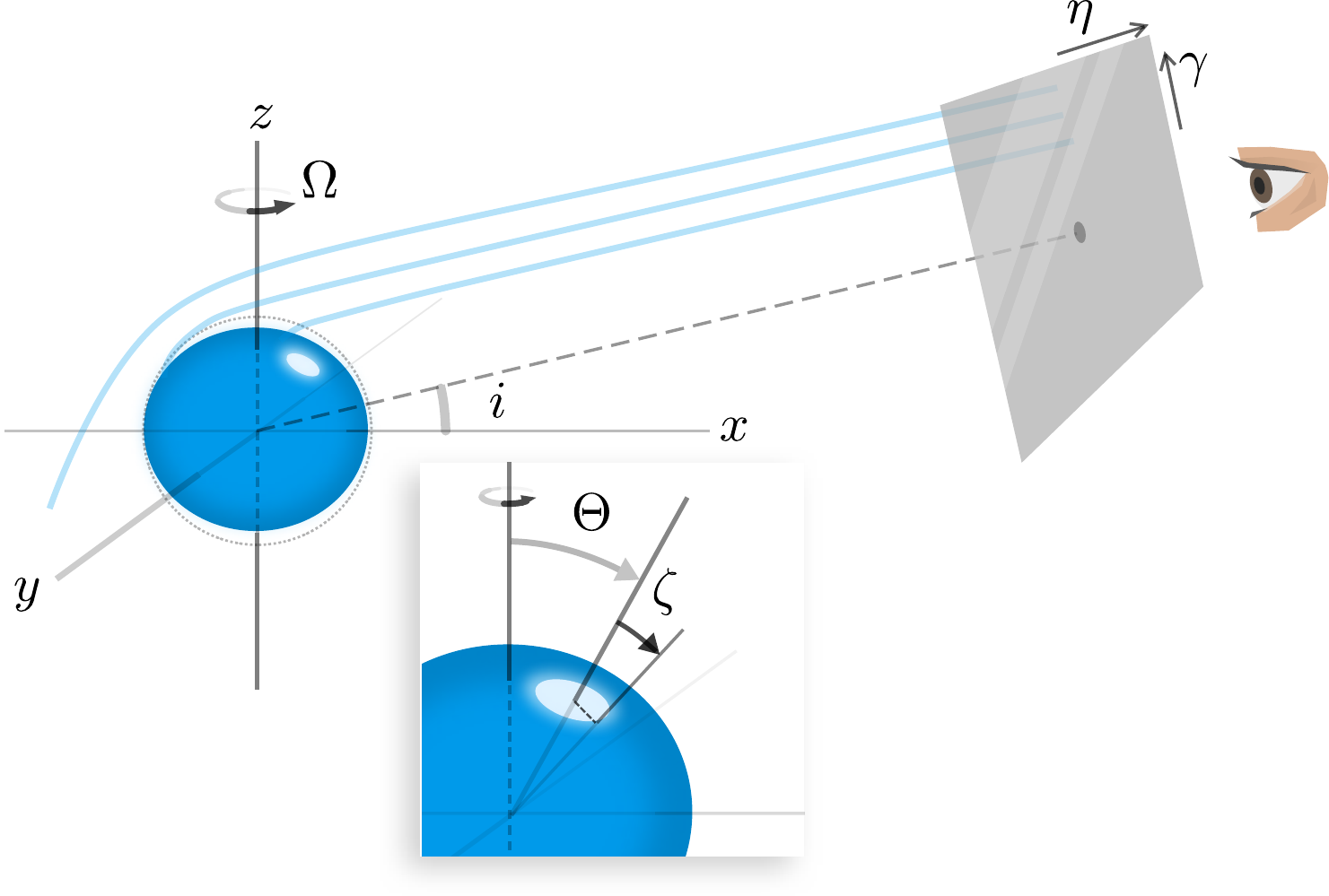}
\caption{Setup for ray tracing. Geodesics are backwards-integrated to the observer, who is located at a long distance from the neutron star (hence, geodesics are parallel upon arrival), at an inclination $i$ with respect to the equatorial plane. The coordinates of the observation plane are $\eta$ and $\gamma$. In all the figures of this paper, the $x$ axis is chosen such that it coincides with the projection onto the equatorial plane of the line defined by the centre of the neutron star and the centre of the observation plane.}
\label{f: illustration-geodesics}
\end{figure}

\skeleton{Setup}
For the rest of this paper, we consider some applications of ray tracing for light emitted at the surface of a neutron star, as measured by a distant observer, for whom the rays arrive parallel (although in the code, we use a backwards-tracing algorithm that inverts this process). In this section, we determine the shape of the neutron star as seen by a distant observer. In Fig. \ref{f: illustration-geodesics}, we describe our setup. The observer is located at an inclination $i$ with respect to the equatorial plane of the neutron star.

\subsection{Method: determination of the surface}

\skeleton{Division of the observer's plane and of the surface}
We divide the observer's plane into a grid of null geodesics as previously discussed in Sect. \ref{s: numerical}, choosing a $1281\times 1281\unit{px}$ grid. Then, we backtrace null geodesics, focusing on the ones that originate from the surface of the neutron star. In a post-processing stage, the surface of the neutron star is also divided in regions of $\Delta \theta = \Delta \phi = 22.5^\circ$, and the positions of the photons on the surface are classified; with the corresponding positions on the observer's plane forming the image.

\skeleton{Surface detection}
We need a clear function for the surface of the neutron star, $r(\theta)$. A straightforward but computationally intensive approach would be a full search and fit from the numerical pressure or enthalpy field yielded by {\tt RNS}, as explained in, e.g., \cite{2020arXiv200805565S}.

After examining the numerical pressure field, we can approximate the surface to an ellipsoid of a certain ratio of polar to equatorial radii ($R_p/R$) \cite[see][for a deeper look into this approximation]{1993ApJS...88..205L}. The equation $r(\theta)$ of an ellipsoid in polar coordinates (with $\theta = 0$ at the $z$ axis) is
\begin{equation}\label{eq: ellipsoid}
\frac{r(\theta)}{R} = \left[1 + \left(\frac{R^2}{R_p^2} - 1\right) \cos^2\theta\right]^{-1/2}.
\end{equation}
For small deformations, $R_p \sim R$ and so, we expand the right hand side so that
\begin{equation}
 \frac{r}{R} = 1 - \left( \frac{R}{R_p} - 1 \right)\left( \frac{R}{R_p} + 1 \right)\frac{\cos^2\theta}{2}.
 \end{equation}
The second term in parenthesis can be approximated to 2. By using trigonometrical relations and considering the series expansion of $1/w:=R_p/R$ around $w=1$, we can put this relation in the form
\begin{equation}\label{eq: surf}
\frac{r}{R} = \sin^2\theta + \frac{R_p}{R} \cos^2\theta,
\end{equation}
which we use here to model the surface, given $R$ and $R_p/R$ from the {\tt RNS} model. Equation \ref{eq: surf} is the equation-of-state dependent form of the AlGendy--Morsink formula (AGM-EOSD)  \citep{2014ApJ...791...78A}. These authors also fitted the formula to several equations of state so that a relation is obtained, in terms of the angular momentum and the compactness of the neutron star only. This approach introduces further uncertainties due to the differences in shape produced by the equations of state, but it allows the bypass of the numerical computation of the interior solution, and it was used by the NICER mission \citep{2019ApJ...887L..21R}. Other quasi-equation-of-state independent formulae have been built for this purpose \citep{2020arXiv200805565S, 2007ApJ...663.1244M}. In the rest of this article, we use both eqs. \eqref{eq: surf} and \eqref{eq: ellipsoid} and explore the differences in the most extreme cases. As the AlGendy--Morsink formula is widely used, we give priority to it in the analysis that follows, clearly marking the figures only if eq. \eqref{eq: ellipsoid} is used instead of \eqref{eq: surf}.

\subsection{Results and comparison of the neutron star shape}
\begin{figure*}
	\includegraphics[width=\textwidth]{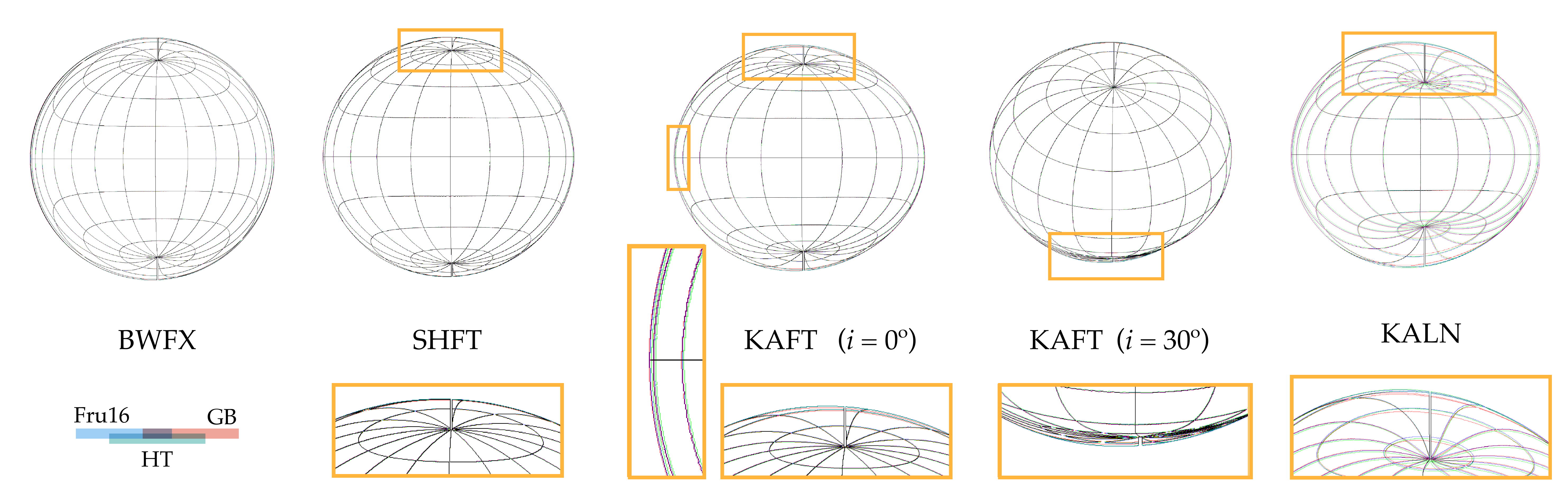}
	\caption{Shape of the neutron star as seen by a distant observer. For {\tt BWFX}, {\tt SHFT} and {\tt KALN}, the inclination angle is $0^\circ$. The parallels and meridians are drawn every $22.5^\circ$. The zoomed-in yellow rectangles show the regions with bigger differences between the metrics. The left side of the neutron star is receding from the observer, while the right side is approaching the observer. The observer's coordinates $(\eta,\gamma)$ have arbitrary scaling, but in this image, they have been normalized such that the \emph{apparent} equatorial radii are the same; flattening can be compared across configurations.}
	\label{f: shape}
\end{figure*}

\skeleton{Parameter study}
Figure \ref{f: shape} shows the surface of the neutron star as seen by a distant observer, for the configurations of Table \ref{t: config}. The shape of the neutron star becomes increasingly flattened with increasing rotation and quadrupole moment. The very high compactness shown in {\tt KALN} greatly increases the area of the far side of the neutron star visible to the observer; this is also visible when comparing configurations {\tt BWFX} and {\tt SHFT}.  By increasing rotation, the asymmetry of the image caused by frame dragging starts to become more apparent.

A caveat of these results is, upon close examination of Fig. \ref{f: shape}, the existence of a missing region, one pixel wide, near the poles. This corresponds to the coordinate singularity of the $z$ axis. Due to the high resolution of the grid used, however, the effects of this singularity are minimized and can be neglected in the applications in the following sections.

\skeleton{Differences between metrics}
The differences between the results yielded by the different metrics are negligible in configuration {\tt BWFX}. They become barely visible in the configuration {\tt SHFT}, i.e., with a pulsar of $\sim 700 \unit{Hz}$. Examining the differences in the image for {\tt KAFT} and {\tt KALN}, we see that the GB metric yields a more oblate object compared to the results yielded by the Fru16 metric. The biggest differences are concentrated in the rays that come to the observer from the edges and far side of the neutron star. For an inclination of $0^\circ$, this includes mainly the polar regions, precisely where the hotspots are expected to be located; for an inclination of $30^\circ$, the differences are concentrated in the rays that come from the southern hemisphere and the top edge of the image, which includes the far side of the polar area. In \cite{2019ApJ...887L..21R} and \cite{Miller_2019}, the models yielded a similar situation: an inclination of $\approx 54^\circ$ with hotspots located in the southern hemisphere.

We calculated the relative differences in the ``apparent area'' or total subtended solid angle of the neutron star as seen by the distant observer. The rays in both the HT and GB metrics produce a slightly smaller total solid angle than the Fru16 metric (for {\tt KAFT}, the differences are of about 0.25 and 0.5 per cent respectively, while for {\tt KALN}, around 0.6 and 1.2 per cent).

\subsection{Effect of the surface formula on the shape}

\begin{figure}
	\includegraphics[width=\columnwidth]{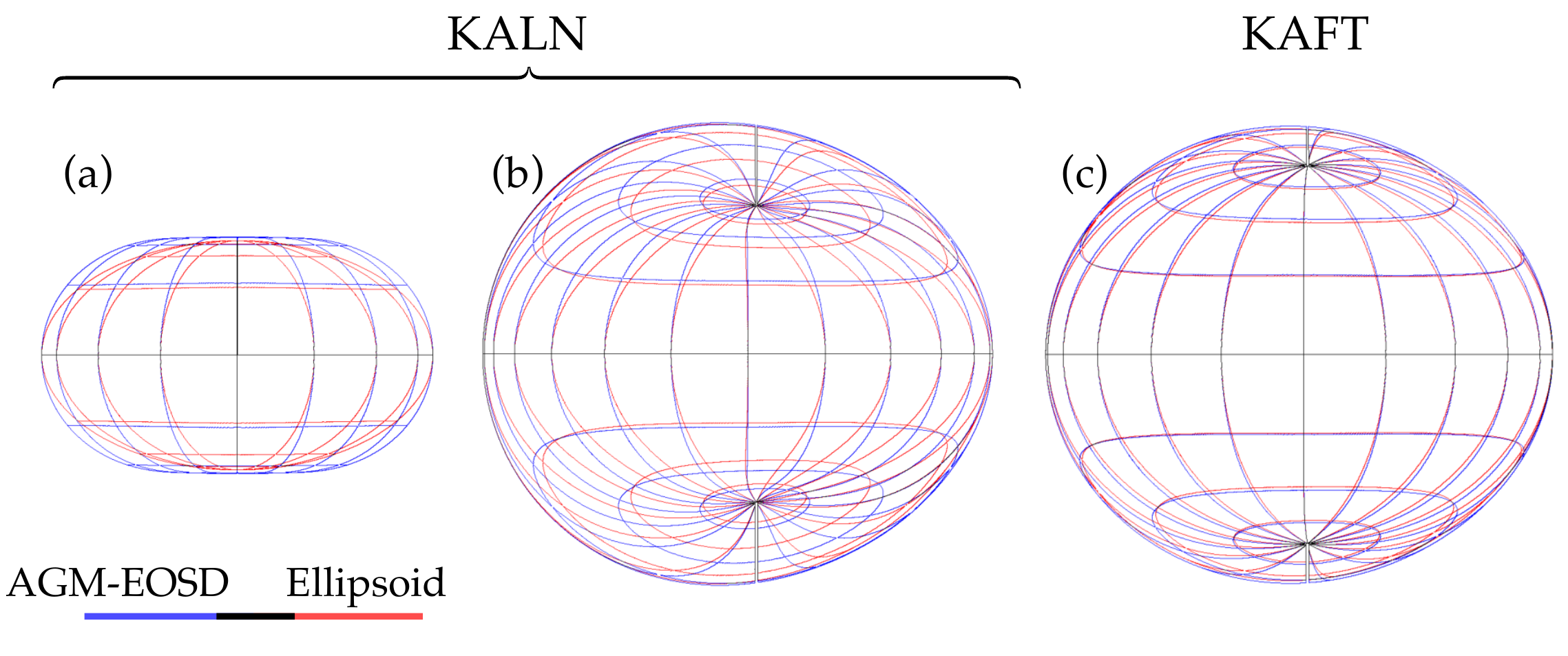}
	\caption{Comparison of surface formulae. For panel \emph{a}, the non-relativistic shape of the neutron star according to the formulae is shown, while panels \emph{b} and \emph{c} use the Fru16 metric.}
	\label{f: surf-formulae}
\end{figure}

In configuration {\tt KALN}, the ratio of polar to equatorial radius is $0.57$, which is already too high for the approximations made in eq. \ref{eq: surf}, and therefore explore the differences between the shapes obtained with eqs. \ref{eq: surf} and \ref{eq: ellipsoid}. The result is shown in Fig. \ref{f: surf-formulae}. For the configuration {\tt KALN}, the differences are appreciable even in the non-relativistic plotting of the surface formulae. For both configurations {\tt KAFT} and {\tt KALN}, however, the differences are appreciable and of the same order as the differences between metrics. The ellipsoid formula has the general effect of increasing the sections of the far side of the neutron star that are visible to the observer, and to a lesser extent, reduce slightly the total solid angle subtended by the neutron star. We refer to the consequences of these effects in the following sections.

\section{Thermal spectrum} \label{s: thermal}
Observed thermal spectra of neutron stars can be used to constrain their masses and radii. In the approximation of a non-spinning, spherical neutron star, the Stefan-Boltzmann law can be used to calculate an apparent radius, given measurements of the bolometric flux and the effective temperature \citep[see, for example, ][]{doi:10.1146/annurev-astro-081915-023322}. Then, the physical radius of the neutron star can be related to the apparent radius by considering lensing in the Schwarzschild spacetime.

In a rapidly rotating neutron star, the effects of rotation and quadrupole moment have to be considered. With high-accuracy knowledge of the radii of the neutron stars \citep[5 to 10 per cent, see][]{2001ApJ...550..426L, 2009PhRvD..80j3003O}, it is possible to constrain the equation of state of the neutron star. In a realistic system, however, the propagation of light through the stellar atmosphere and the environment has to be considered \citep[see][]{2006ApJ...644.1090H}. Additionally, neutron stars with X-ray bursts require more complicated spectra, as discussed in detail in \cite{2020A&A...639A..33S}. In \cite{2015ApJ...799...22B}, rotation and deformation were considered for rapidly-rotating neutron stars ($600\unit{Hz}$) using the GB metric, and corrections were established for non-relativistic fits of the thermal spectrum. We investigate in this section the effects of higher spins and quadrupole moments, and the discrepancies between different approximate metrics, restricting ourselves to the simplest case of pure blackbody emission.

\subsection{Method}

\subsubsection{Gravitational and Doppler shift}
First, the frequency shift due to rotation and the gravitational field has to be calculated, for which we follow \cite{Radosz_2008}, and that we summarize as follows. The frequency $\omega$ of a photon as measured by an observer that moves with a four-velocity $U^\alpha$ is

\begin{equation} \label{eq: grsh1}
	\omega = - k_\alpha U^\alpha = - g_{\alpha\beta}k^\beta U^\alpha.
\end{equation}

A static observer at infinity has a four-velocity $(U_\infty^\alpha) = (U_\infty^t, 0, 0, 0)$. The four-velocity can be normalized with the time-like geodesic condition so that $U_{t,\infty} U ^t_\infty = -1$, yielding

\begin{equation} \label{eq: grsh2}
 U_\infty^t = \sqrt{-1/g_{tt}}.
\end{equation}

An observer located at the surface of the neutron star, co-rotating with it at an angular frequency $\Omega \equiv 2\pi f_\text{rot}$ has a four-velocity $(U^\alpha_\text{surf}) = (U^t_\text{surf}, 0 , 0 , U^\phi_\text{surf})$.  The definition of $\Omega$ also implies that
\begin{multline} \label{eq: grsh3}
\Omega = \left(\frac{d\phi}{dt}\right)_\text{surf} = \left(\frac{d\phi}{d\tau}\frac{d\tau}{dt}\right)_\text{surf} = \frac{U^\phi_\text{surf}}{U^t_\text{surf}}
 \implies U^\phi_\text{surf} = \Omega U_\text{surf}^t,
\end{multline}
where $\tau$ is the proper time measured by the observer. The normalization condition yields, for $U^t_\text{surf}$,
\begin{equation} \label{eq: grsh4}
U^t_\text{surf} = \sqrt{\frac{-1}{g_{tt} + 2g_{t\phi}\Omega + g_{\phi\phi} \Omega^2}}.
\end{equation}

The redshift $\mathbb{Z} = \omega_\text{surf}/\omega_\infty - 1 $ is then calculated for every geodesic that lands on the surface of the neutron star by making use of equations \eqref{eq: grsh1}, \eqref{eq: grsh2}, \eqref{eq: grsh3} and \eqref{eq: grsh4}.

\subsubsection{Specific flux}
Next, following the discussion in \cite{2015ApJ...799...22B}, we calculate the flux measured at infinity $F_\infty$ as
\begin{equation}\label{eq: therm1}
F_\infty(\omega_\infty) = \frac{1}{D^2} \int\int I_\infty (\omega_\infty,\eta,\gamma)\, d\eta\, d\gamma,
\end{equation}
where $D$ is the distance from the observer at infinity to the neutron star. $I_\infty$ is the intensity, that we take as the black body intensity, as seen by an observer at the surface. The quantity $I/\omega^3$ is Lorentz invariant, so,

\begin{equation} \label{eq: therm2}
\frac{I_\infty(\omega_\infty)}{\omega^3_\infty} = \frac{I_\text{surf}(\omega_\text{surf})}{\omega^3_\text{surf}}
\end{equation}
and, defining $f_r := \omega_\infty/\omega_\text{surf}$, we can finally evaluate the flux as

\begin{equation}\label{eq: therm3}
F_\infty (\omega_\infty) = \frac{1}{D^2} \int \int f_r^{3}\, I_\text{surf}\left(\frac{\omega_\infty}{f_r}\right)\, d\eta\, d\gamma.
\end{equation}

The flux reported here has arbitrary units, and so, the choices of the distance to the observer and angular size of the neutron star are also arbitrary.

\subsection{Results of gravitational redshift} \label{s: res-grsh}

\begin{figure*}
	\includegraphics[width=\textwidth]{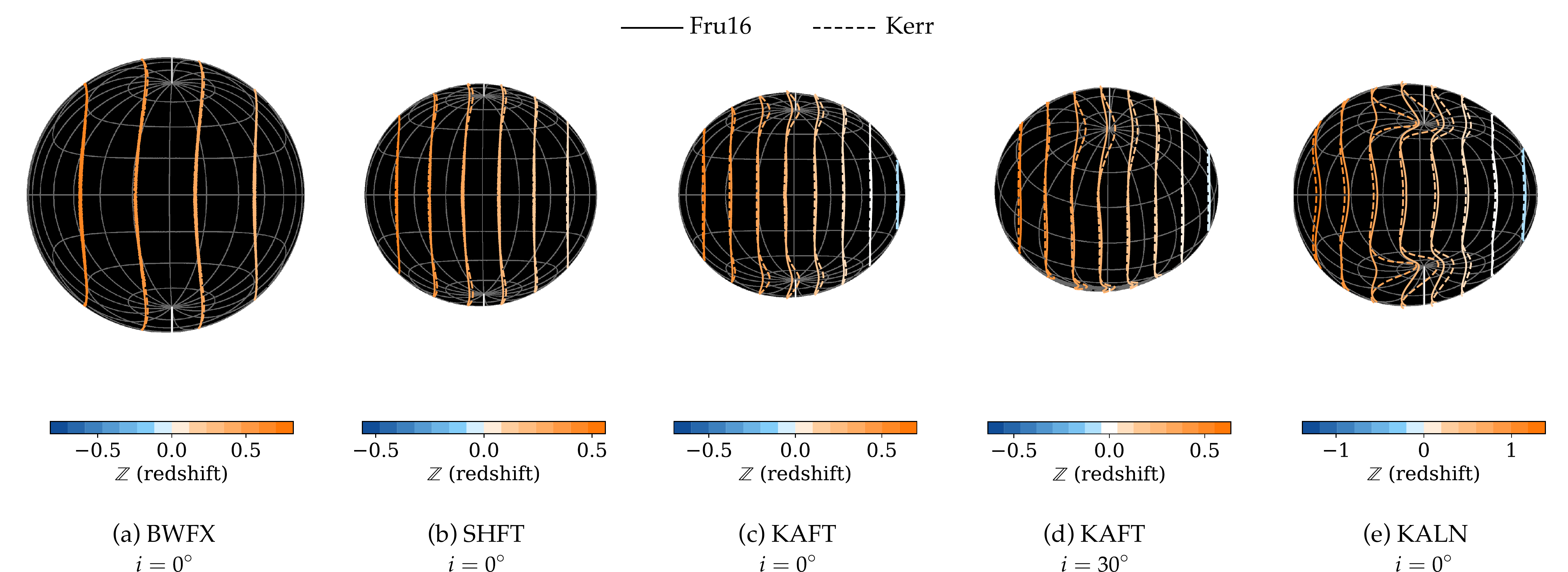}
	\caption{Contours of equal redshift, as computed with the Fru16 and Kerr metrics, showing the effect of the mass quadrupole moment. The observer's coordinates $(\eta,\gamma)$ have arbitrary scaling, but here are normalized such that the neutron star equatorial (physical) radii are the same.}
	\label{f: grsh}
\end{figure*}

\skeleton{GR Doppler shift maps}
In Fig. \ref{f: grsh}, we present the distribution of $\mathbb Z$ on the surface of the neutron star, calculated using the Fru16 and the Kerr metrics. For the configurations {\tt KAFT} and {\tt KALN} there is a net blueshift on the side of the neutron star approaching the observer. With increasing rotation, the Doppler shift due to rotation starts dominating over the purely gravitational redshift (which is axisymmetrical). This is the case despite the higher compactness of {\tt KALN} relative to {\tt KAFT}. The contour lines calculated with the Kerr metric and the oblate surface depart considerably from the Fru16 case around the poles with increasing rotation. In Fig. \ref{f: raytr-aq}b, we showed how the quadrupole moment reduces the deflection angle near the poles, which also has the effect of reducing the redshift.

\subsection{Results and discussion of the thermal spectra}

\begin{figure*}
\setlength{\subfiglen}{0.24\textwidth}

\subfloat[][{\tt BWFX}]{\includegraphics[width=\subfiglen]{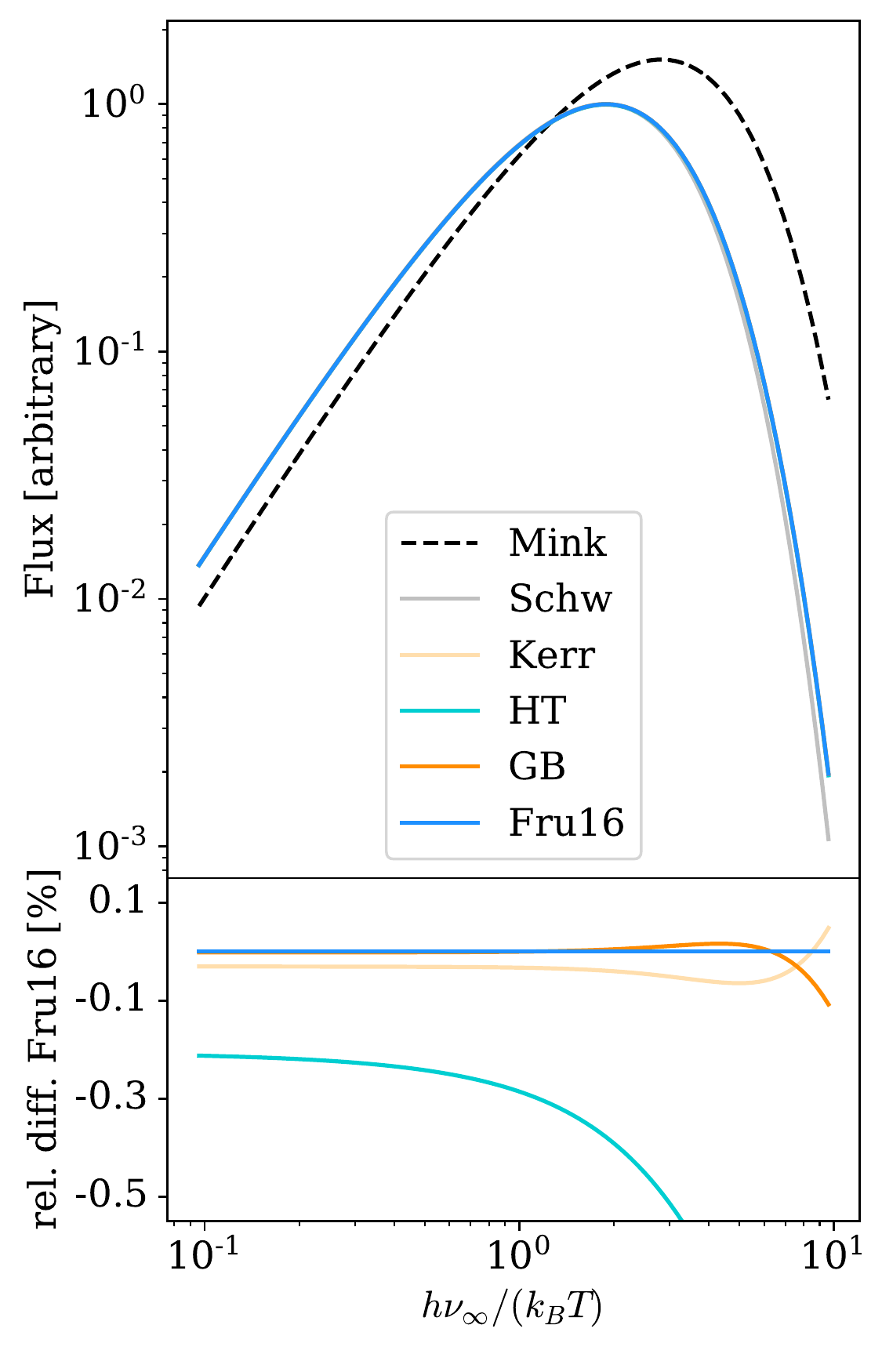}}
\subfloat[][{\tt SHFT}]{\includegraphics[width=\subfiglen]{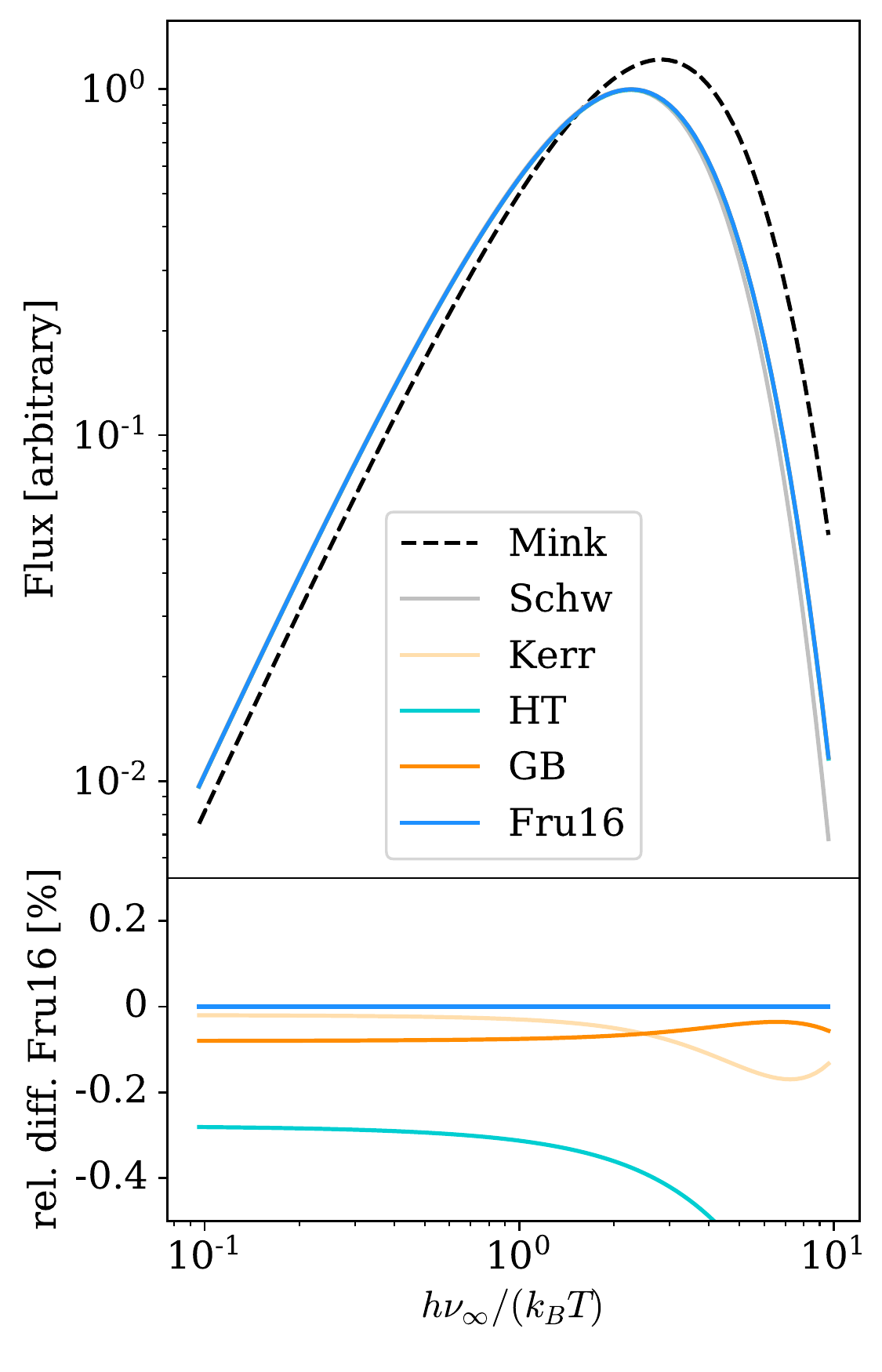}}
\subfloat[][{\tt KAFT}]{\includegraphics[width=\subfiglen]{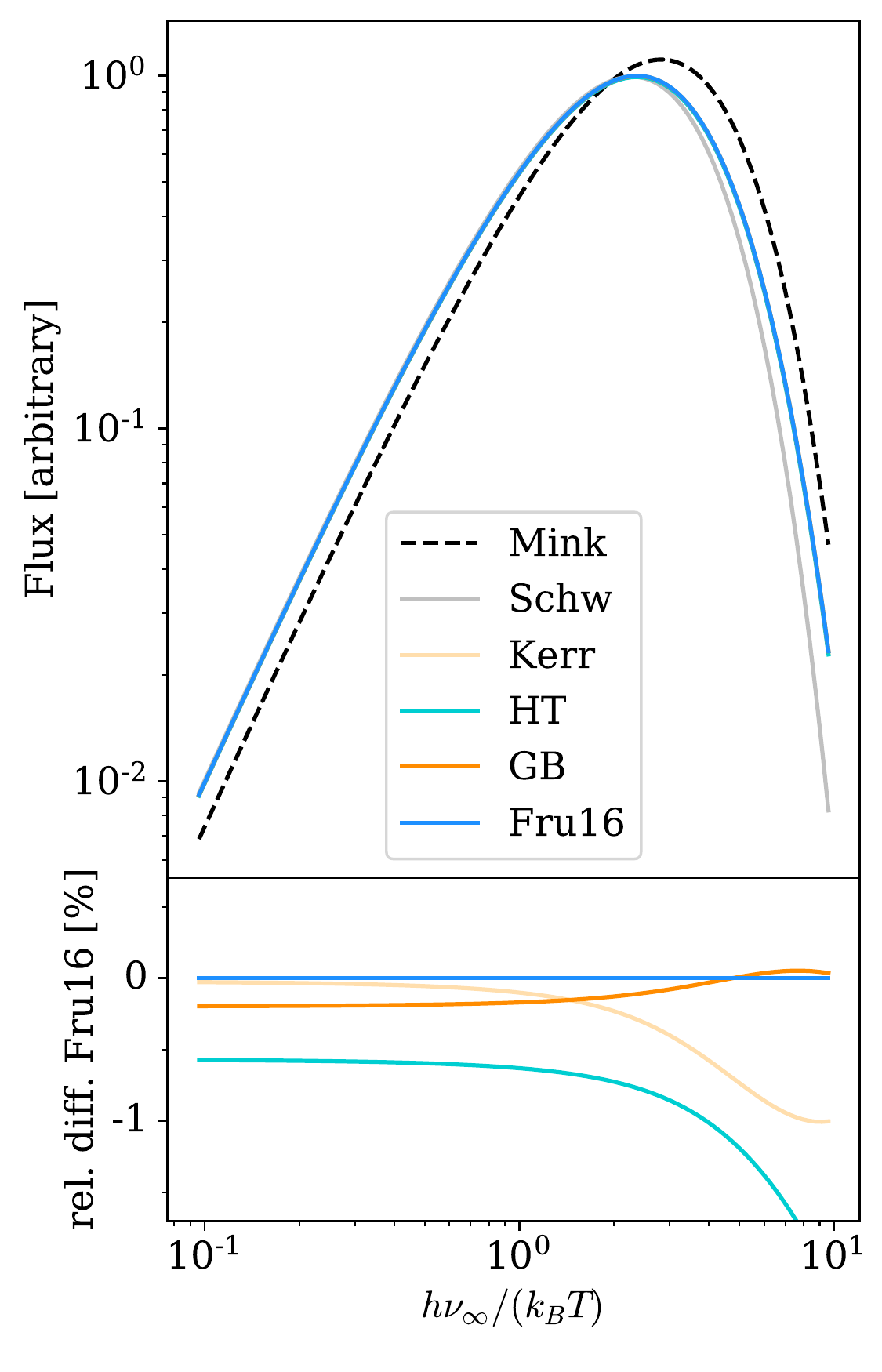}}
\subfloat[][{\tt KALN}]{\includegraphics[width=\subfiglen]{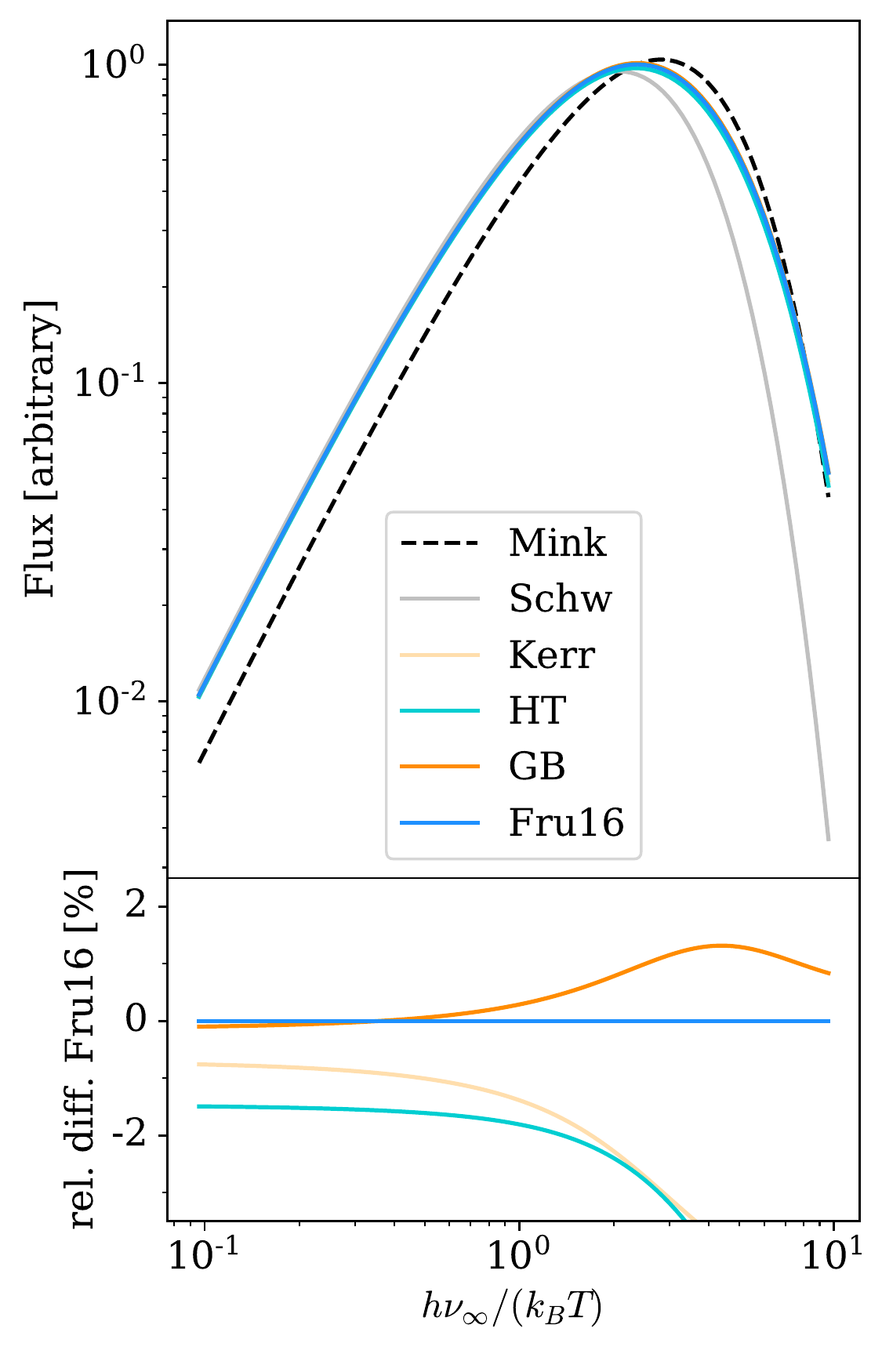}}

	\caption{Thermal (black body) spectrum for different neutron stars. The relative difference with the Fru16 metric, $100\times(F-F_\text{Fru16})/F_\text{Fru16}$,  is shown in the bottom. The fluxes are normalized to the maximum of $F_\text{Fru16}$.}
	\label{f: thermal}
\end{figure*}

\begin{figure}
	\includegraphics[width=\columnwidth]{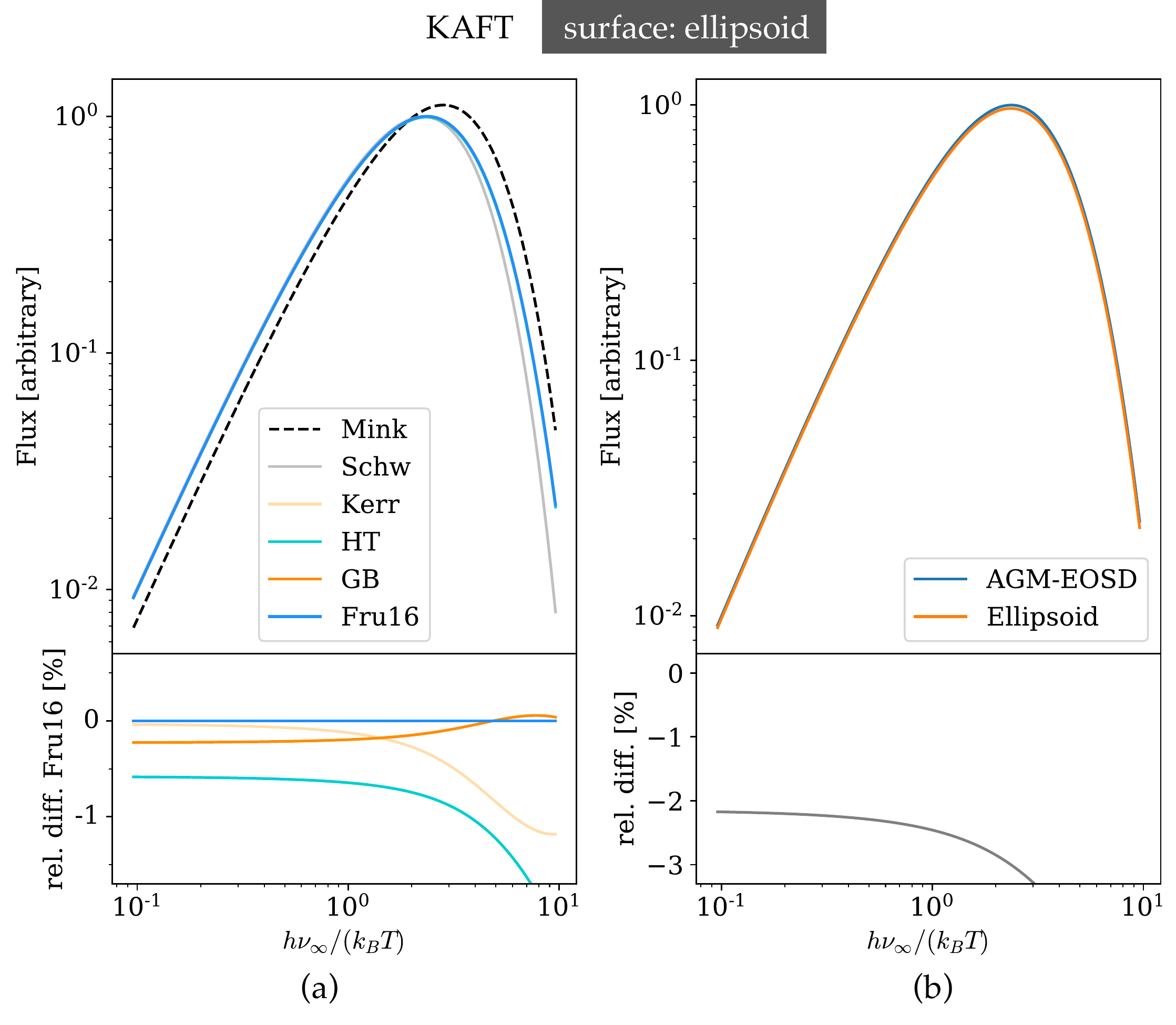}
	\caption{(a) Spectrum computed with the ellipsoid surface formula. (b) Difference between the spectra computed with both surface formulae, using the Fru16 metric.}
	\label{f: therm-ellipsoid}
\end{figure}

\skeleton{General features of the spectra (parameter scan)}
The thermal spectrum of {\tt BWFX} (Fig. \ref{f: thermal}) was built with similar parameters than those used in \cite{2015ApJ...799...22B}. The inclination between the observer and the rotation axis is $0^\circ$ in all cases. The spectrum is simultaneously broadened towards lower frequencies and redshifted due to the increase in apparent area and the presence of gravitational redshift, when comparing it to the (non-relativistic) surface spectrum (here calculated by using the Minkowski metric). When increasing the spin, however, the peak of the spectrum becomes closer to its surface value, although the whole curve continues to be broadened and redshifted. This is a result of the Doppler shift from rotation becoming more dominating compared to the purely gravitational redshift: when the integration over the solid angle is carried out, the blueshifted areas partially compensate the redshift. This means that in principle, the corrections to the peak temperature of a rapidly rotating neutron star become smaller compared to the necessary corrections for moderate rotation.

\skeleton{Differences between metrics}

Each panel in Fig. \ref{f: thermal} shows in the bottom the relative difference between the Kerr, HT and GB metrics, and the Fru16 metric. These differences have their origin in both the total solid angle subtended by the neutron star and the gravitational redshift distribution.

In the case of the configuration {\tt BWFX}, the GB metric and the Fru16 metric are virtually indistinguishable, while the difference in the peak specific flux between those metrics and the expanded HT metric remain below 0.5 per cent. That means that our results agree completely with those presented in \cite{2015ApJ...799...22B}. For the configuration {\tt SHFT}, the differences in the peak between all the metrics account for less than 1 per cent, while in the configurations {\tt KAFT} and {\tt KALN} they are of about 0.5 per cent and 2 per cent, respectively.

Although a full interpretation of the origins of the differences between metrics with quadrupole moment becomes complicated because of the calculations involved, in general terms, the relative differences can be explained as follows. Following eq. \eqref{eq: therm3}, the differences in redshift become increasingly important at high frequencies, where the flux suffers metric-dependent shifts.  For low frequencies, the differences in the total subtended solid angle dominate (which explains the difference of a factor). Since HT produces a smaller total solid angle in comparison to Fru16, the flux is lower as well. The relative differences between the results for the Kerr metric, and GB or Fru16 increase for high frequencies in the most extreme cases. This can be explained by the differences in redshift shown in Fig. \ref{f: grsh}, due to the effect of the quadrupole moment.

\skeleton{Differences between surface formulae}
We also computed the spectrum for configuration {\tt KAFT} using the ellipsoid surface formula. The results are displayed in Fig. \ref{f: therm-ellipsoid}a, and are very similar to Fig. \ref{f: thermal}c. By comparing the curves yielded by both formulae using the Fru16 metric (Fig. \ref{f: therm-ellipsoid}b), a relative difference of about 2 per cent becomes apparent, which means that the use of a realistic surface formula can have a bigger impact than the use of a realistic metric in calculating the thermal spectrum. Again, this difference arises from both the fact that the ellipsoid formula yields an observed smaller solid angle than the one yielded by the AGM-EOSD formula (cf. Fig. \ref{f: surf-formulae}), and that the photons become more redshifted as they travel through regions closer to the origin and therefore with stronger spacetime curvature.

\skeleton{Conclusions}
These results mean that when calculating the effective temperature of a very rapidly rotating neutron star and fitting the spectrum to a black body, apart from the corrections given in \cite{2015ApJ...799...22B}, additional relative differences of a the order of 1 per cent have to be taken into account due to the approximation of spacetime used, and similar, additional values for the use of different surface approximations. The biggest differences between spacetime models appear at high frequencies, where there is a better opportunity to use them for constraining the value of the quadrupole moment only if a realistic surface formula is used.

\section{Pulse profiles} \label{S: ppm}

As stated earlier, pulse profile modelling is an important technique in the determination of the neutron star parameters. In this section, we compute the light curve associated to a hotspot, and study the effects of the neutron star parameters and metrics used for performing the ray tracing. According to the results reported in \cite{2019ApJ...887L..21R} and \cite{Miller_2019}, the shape, number and location of hotspots on the surface of a neutron star can be complex, and it does not necessarily conform to the classical picture of two antipodal sources. For this reason, we concentrate on the signal emitted by individual hotspots.

\subsection{Method: modelling hotspot emission}

\skeleton{Discretization of the observer's plane, characterization of hotspot}
For simplicity, we consider an individual hotspot of a given shape and location on the surface of the neutron star, and that emits monochromatically (so that only for one frequency, we set $I_\text{surf}(\omega_\text{surf}) \equiv \text{const}$ in eq. \eqref{eq: therm3}). We focus on circular hotspots, but we also calculated the pulse profile of a crescent-shaped hot area, such as the ones obtained in the final models of \cite{2019ApJ...887L..21R}. A circular hotspot is characterized by its angular radius $\zeta$ and its colatitude $\Theta$, measured with respect to the rotation axis, as shown in the zoomed-in portion of Fig. \ref{f: illustration-geodesics}.

First, we divide the observer's plane into a grid of pixels as explained in Sect. \ref{s: shape}. We also discretize a full revolution of the neutron star into steps of $\Delta \phi_\text{ns} = 0.5^\circ$.

\skeleton{Time of arrival}
We calculate the time that it takes for photons emitted from the hotspot to reach the observer as
\begin{equation}
 t_\text{arrival} = t_\text{grav} + t_\text{rot}.
 \end{equation}
Here, $t_\text{grav}$ is the travel time between the surface of the neutron star and the observer, and $t_\text{rot} = \phi_h/(2\pi f)$ is the time at which the photons are emitted as the hotspot rotates; $\phi_h$ is the azimuth of the center of the hotspot measured with respect to the $x$ axis (i.e., the axis that coincides with the normal to the observer's plane when the inclination is zero). For simplicity, and given that the observer is located at an arbitrary distance from the hotspot, we use the time of arrival of the photon that lands in the center of the observer's plane when $\phi_h = 0$ as a reference time, so that we henceforth use $\Delta t_\text{arrival}$, the difference in time of arrival. We calculate $\Delta t_\text{arrival}$ for each discretized azimuthal position, and for every point of the hotspot that reaches the observer's discretized plane.

\skeleton{Collection of photons, image of the hotspot, light curve}
Then, we collect all photons that reach the observer's plane in an interval $\Delta t_\text{expos} = \Delta \phi_\text{ns}/(2\pi f)$, in analogy to the exposure time of a detector located there. These photons form the image of the hotspot. With the image of the hotspot, we calculate the total flux emitted from the hotspot, as done in Sect. \ref{s: thermal}, as a function of time (light curves).

\skeleton{Characterization of crescent-shaped hotspots}
\cite{2019ApJ...887L..21R} and \cite{Miller_2019} investigated not only circular hotspots, but also configurations of higher complexity, like crescent-shaped hotspots. \cite{Miller_2019} also considered oval hotspots with allowed overlapping.

We decided to take the same parameters as reported in \cite{2019ApJ...887L..21R} as a reference to study the light curve of a crescent-shaped hotspot of a kilohertz pulsar. In this case, the crescent is formed by subtracting two circular-shaped hotspots, the superseding member (subscript $s$) and the ceding member (subscript $c$). An additional parameter in this case is the difference in azimuthal position between the two members.

\subsection{Results and comparison of the pulse shapes}

\begin{figure*}
	\includegraphics[width=0.9\textwidth]{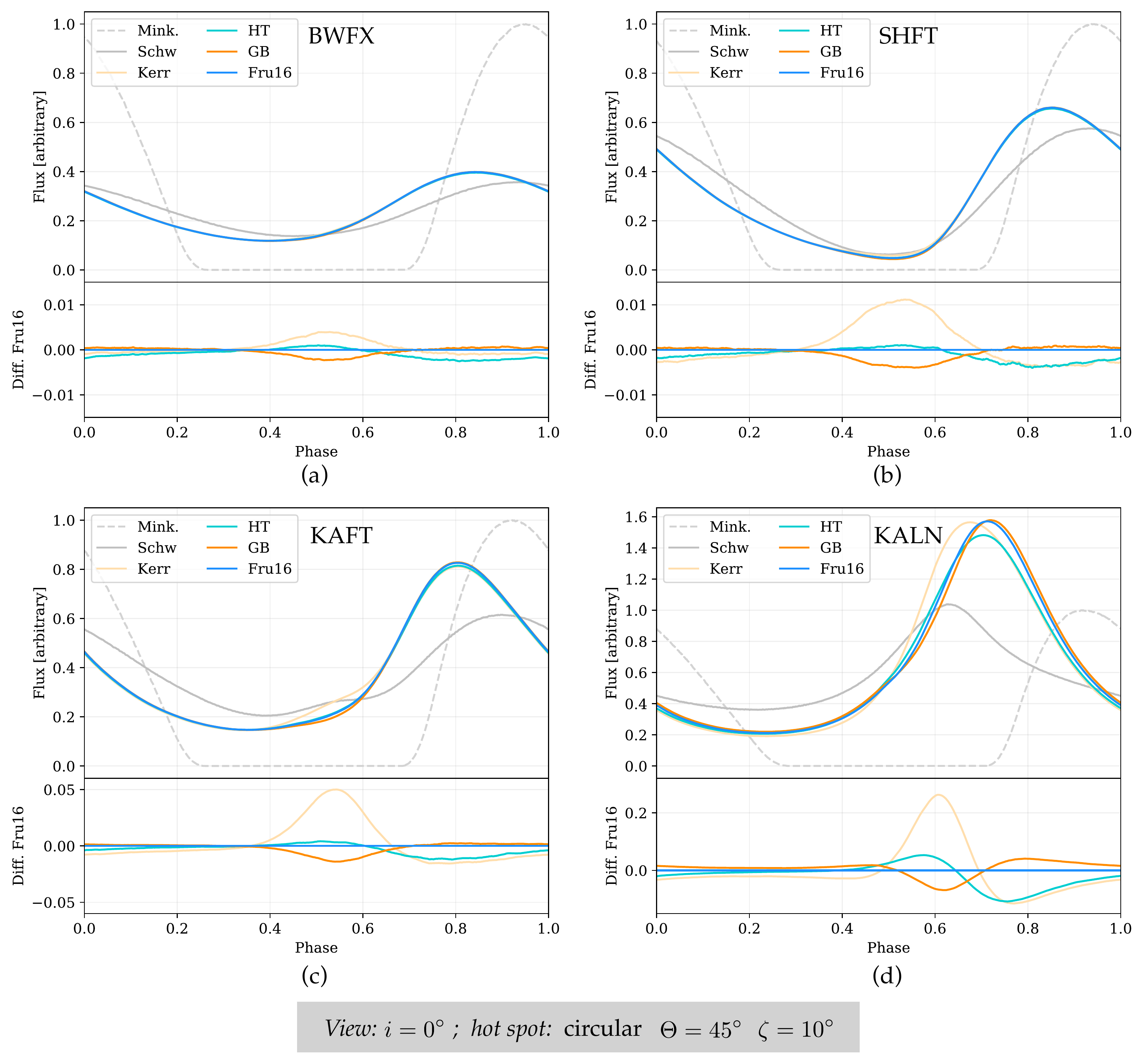}
	\caption{Light curves from hotspots: comparison of the neutron star configurations and metrics.}
	\label{f: hotspot-configs}
\end{figure*}

\begin{figure*}
	\includegraphics[width=0.9\textwidth]{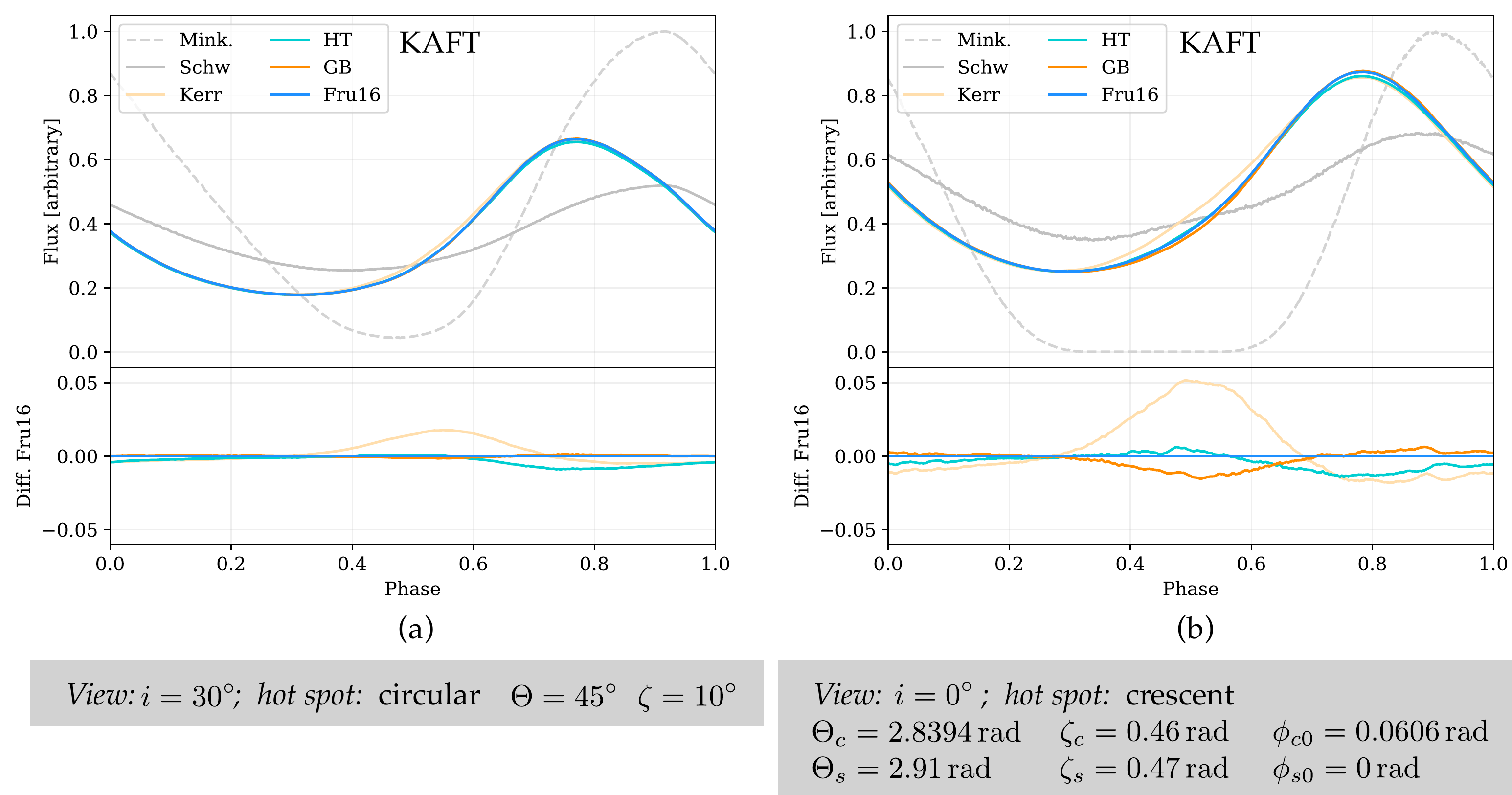}
	\caption{Light curves from hotspots: effect of the inclination with respect to the observer.}
	\label{f: hotspot-incl}
\end{figure*}

\subsubsection{Effects of the neutron star and hotspot configurations}

\skeleton{Effects of the configuration}
Figure \ref{f: hotspot-configs} shows the light curves for a circular hotspot located at $\Theta = 45^\circ$, and an angular radius of $\zeta = 10^\circ$, viewed from an inclination of $i = 0^\circ$. The gravitational lensing of the surface causes the pulse to be visible at all times, in contrast to the dashed curve (that goes to zero), which corresponds to a non-relativistic treatment, and was calculated by ray-tracing with the Minkowski metric.

With increasing rotation and quadrupole moment, there are several clearly visible effects. A phase shift with respect to the non-relativistic light curve is due to the differences in time of arrival of the signal caused by the curvature of spacetime (including frame dragging). The rising of the peak of the pulse compared to the non-rotating (Schwarzschild) case is product of the Doppler boost caused by rotation. These two effects were also seen in \cite{2006MNRAS.373..836P} and \cite{2014ApJ...792...87P}. In configurations {\tt KAFT} and {\tt KALN}, the phase shift (with respect to Minkowski) is specially visible: the pulse is already descending when the Minkowski counterpart is ascending. The Doppler boost is increasingly more apparent with rotation when comparing the peaks of the profile across configurations. Additionally, we see that the flux is overall lower than its non-relativistic counterpart for configuration {\tt BWFX}, and increasingly higher with rotation and quadrupole moment in the other configurations. This is simply due to the fact that a flat spacetime produces no redshift, and so, $f_r = 1$ in \eqref{eq: therm3}, whereas for curved spacetimes, $f_r < 1$. High compactness and high mass quadrupole moment increases the apparent area of the hotspot due to gravitational lensing and flattening. This can be seen when comparing the general relativistic profiles of configurations {\tt KAFT} and {\tt KALN} with their respective non-relativistic counterparts, as well as comparing the areas of the surface grids for the same configurations in Fig. \ref{f: grsh}.

\skeleton{Effects of the size of the hotspot}
We also tried varying the angular radius of the hotspot, with values of $5^\circ$, $10^\circ$ and $15^\circ$, finding essentially no difference between the profiles except for a different normalization due to the variation in area. These results are not shown in our figures, but are in agreement with the findings of \cite{Baub_ck_2015}, who found that hotspots with angular radii of $\lesssim 18^\circ$ yield very similar profiles, leading to errors of $\lesssim 10$ per cent when used for determining the radius of the neutron star.

\subsubsection{Effects of the use of different metrics}

\skeleton{Differences between metrics}

As with the other applications presented in this paper, the differences between metrics are only present in the configurations with higher spin. The lower panels in for each configuration plot in Fig. \ref{f: hotspot-configs} show the absolute difference between each metric and the Fru16 metric. The presence of a grid introduces resolution-dependent noise in the curve, and so, a moving average and its corresponding standard deviation were calculated to show the overall behaviour of the difference (the flux is not averaged). For the configuration {\tt BWFX}, the differences between metrics are negligible. For the configurations {\tt KAFT} and {\tt KALN}, the differences shown are bigger than their standard deviation, and for {\tt SHFT}, they are of the same order of magnitude.

\skeleton{Effects of the location of the hotspot}
These differences between metrics are caused by differences in redshift, time of arrival and shape of the neutron star, which we discussed in the preceding sections. Therefore, hotspots that move close to the edge of the image as seen by the observer produce the biggest differences on the profile. In particular, the hotspot configuration selected in Fig. \ref{f: hotspot-configs} ($\Theta=45^\circ$, $\zeta=10^\circ$) with an inclination of $i=0^\circ$, goes into an area of high difference. By comparing the result yielded by the Kerr metric and the HT, GB and Fru16 metrics, we determine that the difference arises due to the presence of the quadrupole moment. At that point, the hotspot is at the far side of the neutron star as seen by the observer, and close to the edge of the image, where the photons arriving to the observer are emitted more tangent to the surface, spend more time in the strong gravitational field and therefore are also deflected the most. Hotspots that move into those regions are therefore more likely to be strongly influenced by the quadrupole moment. In Fig. \ref{f: hotspot-incl}a, a hotspot at same location is shown, but with an inclination of $i=30^\circ$. As the hotspot moves into the far side of the neutron star with respect to the observer, the same differences arise, but upon comparison with Fig. \ref{f: hotspot-configs}c, they are less important, because the hotspot is farther away from the edge of the image.

Apart from the differences close to the valley of the profile, there are some differences at the peak of the profile, but only significant in the most extreme configurations. These differences arise due to the differences in redshift as the quadrupole moment becomes increasingly important. In that region, the results from GB and Fru16 agree, while HT and Kerr underestimate the peak of the profile.

\subsubsection{Crescent-shaped hotspots}
Fig. \ref{f: hotspot-incl}b presents the results for a crescent shaped hotspot, observed with an inclination of $i=0^\circ$. The shape of the pulse changes slightly (see, e. g., the rounding of the valley for the Minkowski case, compared to Fig. \ref{f: hotspot-configs}c). The differences are similar for the region discussed in the previous paragraphs, but the narrower shape of the crescent increases the importance of the time of arrival and redshift (as a narrower range of frequencies and source locations are integrated at a time), resulting in the increase of the differences between metrics also in the peak of the profile.

\subsubsection{Comparison of surface formulae}

\begin{figure*}
	\includegraphics[width=0.9\textwidth]{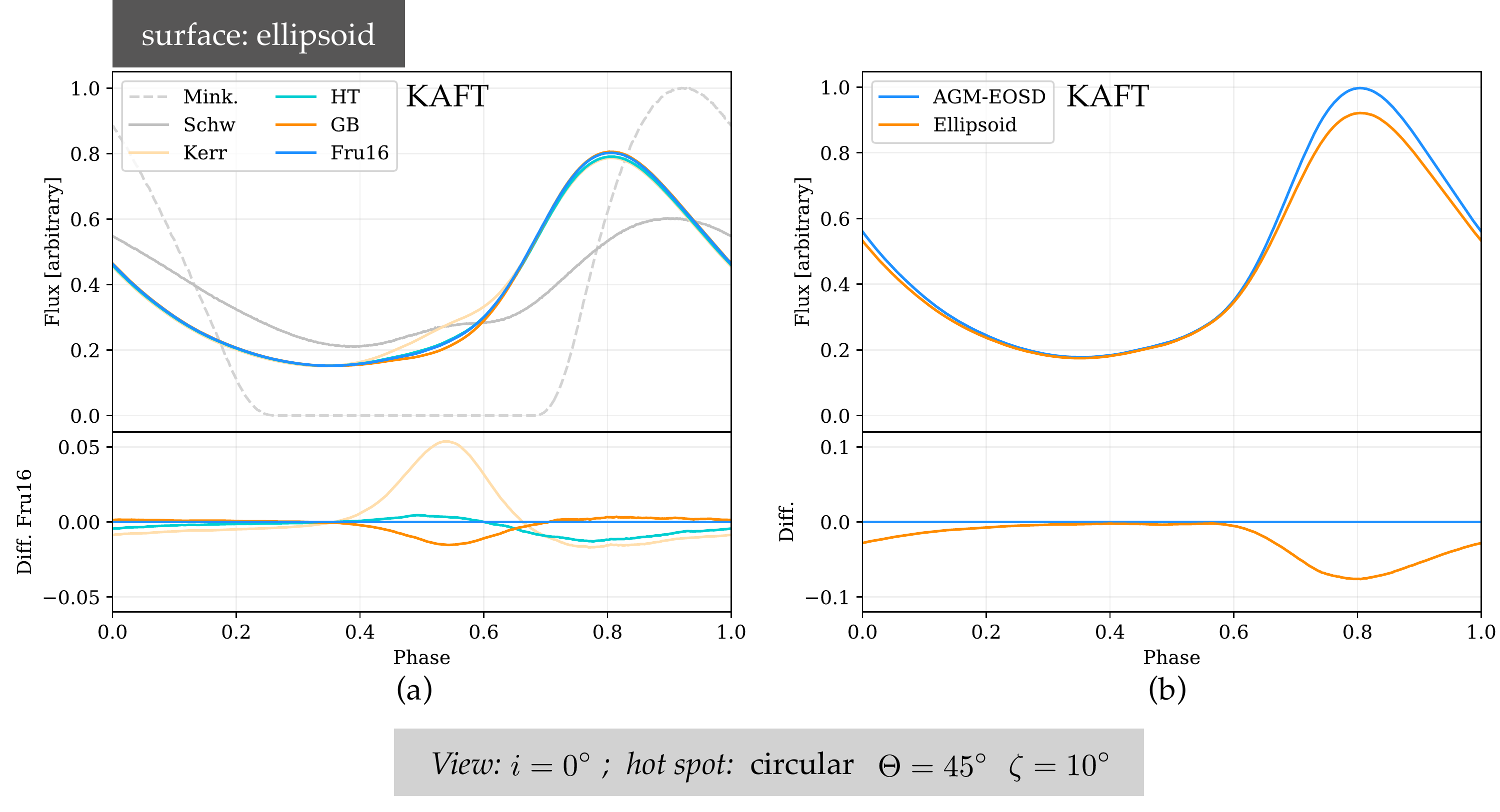}
	\caption{(a) Light curves computed with the ellipsoid surface formula. (b) Differences between formulae, for the Fru16 metric.}
	\label{f: hotspot-ellipsoid}
\end{figure*}

We also include the calculation of a profile for configuration {\tt KALN} using the ellipsoid surface formula, the result of which is shown in Figure \ref{f: hotspot-ellipsoid}a. The differences between the results from the metrics increase slightly (the departure from the Kerr result is specially notable), but the general features discussed in the previous paragraphs are preserved. Figure \ref{f: hotspot-ellipsoid}b shows the differences between the results yielded by the two surface formulae. There are marked differences at the peak of the profile, bigger than the differences between metrics at the peak, a result that highlights the need for choosing a realistic surface formula for modeling pulses under these circumstances. However, the pulse shape remains unchanged at the valley, which means that the differences caused by the treatment of the quadrupole moment between the different metrics are significant, independently from the surface formula used. Theoretically, this particular hotspot configuration could serve as a way of constraining the mass quadrupole moment or a way to probe the accuracy of approximations to the spacetime around a neutron star.

\subsubsection{Conclusions}
From the preceding discussion, we conclude that in the case of observationally-known pulsars (rotation frequencies $\lesssim 700\unit{Hz}$), the expanded HT metric yields sufficiently accurate pulse shapes. This is a consequence of the findings we discussed in Sect. \ref{s: raytr-results}, namely, that the HT metric yields accurate results if both $a$ and $q$ increase. The quadrupole moment is, however, dependent on the equation of state (we only examined two cases) and other factors like the internal magnetic field pressure. Studies that consider a wide range of values of $a$ and $q$ should use a more accurate approximation than HT. For rotation frequencies $\gtrsim 700 \unit{Hz}$ also more accurate metrics than HT are needed. As discussed in Sect. \ref{S: raytracing}, the results of GB and Fru16 shown here can be used to estimate the results that the exact solutions would yield without the increased computational cost. We showed that the differences between the results yielded by the metrics are relatively small but significant, and occur mostly when the hotspot is located near the edge of the neutron star image as seen by the observer (close to the valley of the light curve), but also at the peak, due to the differences in redshift. We also showed that surface formulae can also yield significant differences, more important when closer to the peak.

\section{Summary and conclusions} \label{s: summary}

We have compared a catalog of different metrics that describe the exterior spacetime around a neutron star, including the numerical solution provided by the code {\tt RNS}. For parameter ranges of mass, specific angular momentum and quadrupole moments that correspond to rapidly rotating neutron stars of rotation frequencies of $\lesssim 1\unit{kHz}$, we found that:

\begin{itemize}
\item All the metrics are consistent with the numerical solution, but we could not use it to discriminate between them.
\item The GB and Fru16 metrics provide excellent approximations compared to the exact solutions by Quevedo--Mashhoon and Manko--Novikov for ray tracing, although the expanded Hartle--Thorne metric provides a reasonable approximation and the lowest computing time.
\item The biggest differences between the results yielded by the metrics are found in the more extreme rotating configurations, and for regions closer to the edge of the neutron star as seen by the observer.
\item The differences between metrics start to be relevant at rotational frequencies of $\sim 700\unit{Hz}$, they are negligible for frequencies of $\sim 600 \unit{Hz}$ (which means that our results agree with previous studies) and they become more important at around $\sim 1000\unit{Hz}$.
\item Due to the dominance of the rotational Doppler shift, the corrections to the thermal spectrum of a rapidly rotating neutron star get smaller than in the slow rotation case, which is dominated by pure gravitational redshift.
\item The discrepancies between metrics in the determination of the peak of the thermal spectrum of a neutron star are of the order of 1 per cent for rapid rotation.
\item For pulsar profile modelling of observable pulsars (rotation frequencies of $\lesssim 700\unit{Hz}$), the expanded Hartle--Thorne metric yields sufficiently accurate pulse shapes when compared to higher-order approximations, while keeping the shortest computing time.
\item Hotspots that go through regions near the visible edges of the neutron star (corresponding to the valley of the pulse) are the most affected by the treatment of the mass quadrupole moment, and therefore could theoretically be used to constrain its value or discriminate between spacetime models.

\end{itemize}

\section*{Acknowledgements}
G.A.O.-M. wishes to thank all the members of the Theoretische Astrophysik group of the University of Tübingen, in particular to Sourabh Nampalliwar, Arthur Suvorov, Daniela Doneva and Kostas Kokkotas, for their valuable suggestions for improving the ray-tracing applications. We wish to thank Javier Bonatti-González for the ideas and input during the development of this work, and the anonymous referee for their valuable comments and suggestions.

\section*{Data availability}
The data analysed in this article can be generated by means of the software package {\tt Ujti} that we developed, the source code of which is publicly available at \url{http://cinespa.ucr.ac.cr}.


\bibliographystyle{mnras}
\bibliography{bibliography}


\appendix
\section{Exterior spacetimes} \label{s: appendix-metrics}

\subsection{Quevedo--Mashhoon (QM) metric} \label{a: QM}
We present here the version of the Quevedo--Mashhoon metric with the parameters written as the mass ($ M $), rotation ($ J = M a $), and mass quadru\-po\-le parameter
\[ q_\text{{\sc qm}} :=  -\frac{15q}{2(\sigma^2/M^2)^{3/2}M^3}. \]
This spacetime in prolate coordinates $ (t, \, x, \, y, \phi) $ is given by

\begin{eqnarray}
\label{QM}
ds^2 &=& - f(d t - \omega d \phi)^2 \\
&+& \frac{\sigma^2}{f}
\left[{\rm e}^{2 \gamma} (x^2 - y^2) \left(\frac{d x^2}{x^2 - 1}
+ \frac{d y^2}{1 - y^2} \right) \right. \\
&+& \left. (x^2 - 1)(1 - y^2) d \phi^2 \right]
\nonumber \\
&=& - V d t^2 + 2 W d t d \phi + X d x^2 + Y d y^2 + Z d \phi^2 ,
\nonumber
\end{eqnarray}

\noindent
where

\begin{eqnarray}
\label{qm1}
f &=& \frac{{\cal{R}}}{{\cal{L}}} e^{- 2 \psi} , \nonumber \\
\omega &=& - 2 \left(a
+ \sigma \frac{{\cal{M}}}{{\cal{R}}} e^{2 \psi} \right), \nonumber \\
{\rm e}^{2 \gamma} & = & \frac{1}{4} \left(1 + \frac{M}{\sigma} \right)^{2}
\frac{{\cal{R}}}{(x^2 - 1)} e^{2 {\chi}} , \nonumber \\
{\cal{R}} &=& a_{+} a_{-} + b_{+} b_{-} , \nonumber \\
{\cal{L}} &=& a_{+}^{2} + b_{+}^{2} , \nonumber \\
{\cal{M}} &=& \left[x (1 - y^{2})(\lambda + \eta) a_{+}
+ y (x^{2} - 1)(1 - \lambda \eta) b_{+} \right] , \nonumber \\
{\psi} &=& q_\text{{\sc qm}} P_{2} Q_{2} , \\
{\chi} &=& \frac{1}{2}(1 + q_\text{{\sc qm}})^{2}
\ln{\left[\frac{x^{2} - 1}{x^{2} - y^{2}} \right]} \nonumber \\
&+& 2 q_\text{{\sc qm}} (1 - P_{2}) Q_{1}
+ q_\text{{\sc qm}}^{2} (1 - P_{2}) [(1 + P_{2})(Q_{1}^{2} - Q_{2}^{2}) \nonumber \\
&+& \frac{1}{2}(x^{2} - 1)(2 Q_{2}^{2}-3 x Q_{1} Q_{2}
+ 3 Q_{0} Q_{2} - Q_{2}^{\prime})] , \nonumber
\end{eqnarray}

\noindent
with

\begin{eqnarray}
\label{qm2}
a_{\pm} &=& x (1 - \lambda \eta) \pm (1 + \lambda \eta) , \nonumber \\
b_{\pm} &=& y (\lambda + \eta) \mp (\lambda - \eta) , \nonumber \\
\lambda &=& \alpha {\rm e}^{2 \delta_{+}} , \\
\eta &=& \alpha {\rm e}^{2 \delta_{-}} , \nonumber \\
\delta_{\pm} &=& \frac{q_\text{{\sc qm}}}{2} \ln{\left[\frac{(x \pm y)^{2}}{x^{2} - 1} \right]}
+ q_\text{{\sc qm}} (P_2 - P_0) Q_1 \pm q_\text{{\sc qm}} P_1 (Q_2 - Q_0) . \nonumber
\end{eqnarray}

\noindent
The functions $ P_l(y) $ and $ Q_l(x) $ are Legendre polynomials of the first and second kind, respectively. They are given by

\begin{eqnarray}
\label{qm3}
P_0 &=& 1 , \nonumber \\
P_1 &=& y , \nonumber \\
P_2 &=& \frac{1}{2} (3 y^2 - 1) , \nonumber \\
Q_0 &=& \frac{1}{2} \ln{\left[\frac{x + 1}{x - 1} \right]} , \\
Q_1 &=& \frac{x}{2} \ln{\left[\frac{x + 1}{x - 1} \right]} - 1 ,
\nonumber \\
Q_2 &=& \frac{1}{4} (3 x^2 - 1) \ln{\left[\frac{x + 1}{x - 1} \right]}
- \frac{3 x}{2} , \nonumber \\
Q_{2}^{\prime} &=& \frac{d Q_{2}}{d x} \nonumber
\end{eqnarray}

\noindent
\noindent
The metric potentials are given by

\begin{eqnarray}
\label{qm4}
V &=& f \nonumber \\
W &=& f \omega \nonumber \\
X &=& \sigma^2 \frac{{\rm e}^{2 \gamma}}{f}
\left(\frac{x^2 - y^2}{x^2 - 1} \right)  \\
Y &=& \sigma^2 \frac{{\rm e}^{2 \gamma}}{f}
\left(\frac{x^2 - y^2}{1 - y^2} \right) \nonumber \\
Z &=& \frac{\sigma^2}{f} (x^2 - 1)(1 - y^2) - f \omega^2 . \nonumber
\end{eqnarray}

\noindent
The mapping to transform from prolate coordinates to spherical coordinates is

\begin{equation}
\label{qm5}
\sigma x = {(r - M)} , \qquad y = \cos{\theta} .
\end{equation}

\noindent
The constants $ \alpha $ and $ \sigma $ are related with $ a $, and $ M $ by means of

\begin{equation}
\label{qm6}
\alpha a = {\sigma - M} , \qquad \sigma = \sqrt{M^{2} - a^{2}} .
\end{equation}

The relativistic multipoles of this metric are

\begin{eqnarray}
\label{qm7}
M_0 &=& M , \nonumber \\
S_1 &=& J = M a , \\
M_2 &=& q_\text{{\sc qm}} \left(1 - \frac{a^{2}}{M^{2}} \right)^{3/2} - M a^{2} . \nonumber
\end{eqnarray}

\subsection{Manko--Novikov (MN) metric} \label{a: MN}

This spacetime in prolate coordinates $ (t, \, x, \, y, \phi) $ has the following potentials:

\begin{eqnarray}
\label{mn1}
f &=& \frac{A}{B} {\rm e}^{2 \psi} , \nonumber \\
\omega &=& - 2 k \left(\frac{2 \alpha}{1 - \alpha^2}
- \frac{C}{A} {\rm e}^{- 2 \psi} \right) ,  \nonumber \\
{\rm e}^{2 \gamma} & = &
\frac{A {\rm e}^{2 {\chi}}}{(1 - \alpha^2)^2 (x^2 - 1)} , \nonumber \\
A &=& (x^2 - 1)(1 + a b)^2 - (1 - y^2)(b - a)^2 , \nonumber \\
B &=& [x + 1 + (x - 1) a b]^2 + [(1 + y) a + (1 - y) b]^2 , \nonumber \\
C &=& (x^2 - 1)(1 + a b) [b - a - y (a + b)] \nonumber \\
&+& (1 - y^2)(b - a) [1 + ab + x (1 - a b)] , \nonumber \\
\psi &=& q_\text{{\sc mn}} \frac{P_2}{R^3} , \nonumber \\
\chi &=& \frac{1}{2} \ln{\left[\frac{x^2 - 1}{x^2 - y^2} \right]}
+ \frac{9}{6} \frac{q_\text{{\sc mn}}^2}{R^6} (P_3 P_3 - P_2 P_2) \\
&+& 2 q_\text{{\sc mn}} \left[x \frac{P_0}{R} - y \frac{P_1}{R^{2}}
+ x \frac{P_2}{R^{3}} - 1 \right] , \nonumber \\
a &=& a(x, \, y) = - \alpha {\rm e}^{- 2 q_\text{{\sc mn}} \chi_1} , \nonumber \\
b &=& b(x, \, y) = \alpha {\rm e}^{2 q_\text{{\sc mn}} \chi_2} , \nonumber \\
\chi_1 &=& - 1 + \frac{(x - y)}{R} \left(P_0 + \frac{P_1}{R}
+ \frac{P_2}{R^{2}} \right) , \nonumber \\
\chi_2 &=& 1 - \frac{(x + y)}{R} \left(P_0 - \frac{P_1}{R}
+ \frac{P_2}{R^{2}} \right) , \nonumber \\
R &=& \sqrt{x^2 + y^2 - 1} , \nonumber \\
P_{n} &=& P_{n} \left(\frac{x y}{R} \right) , \nonumber
\end{eqnarray}

\noindent
with $ P_n $ as the Legendre polynomials.

\noindent
The metric potentials are given by

\begin{eqnarray}
\label{mn2}
V &=& f \nonumber \\
W &=& f \omega \nonumber \\
X &=& k^2 \, \frac{{\rm e}^{2 \gamma}}{f} \left(\frac{x^2 - y^2}{x^2 - 1} \right) \\
Y &=& k^2 \, \frac{{\rm e}^{2 \gamma}}{f}
\left(\frac{x^2 - y^2}{1 - y^2} \right) \nonumber \\
Z &=& \frac{k^2}{f} (x^2 - 1)(1 - y^2) - f \omega^2 . \nonumber
\end{eqnarray}

\noindent
The mapping to transform from prolate coordinates to spherical coordinates is the same as in (\ref{qm5}), and the constants $ \alpha $ and $ k $ are related with $ a $, and $ M $ by means of the same expressions as in (\ref{qm6}), because $ k \equiv \sigma $ in the QM metric.

\noindent
The first three relativistic multipoles are given by

\begin{eqnarray}
\label{mn3}
M_0 &=& M = k \left(\frac{1 + \alpha^{2}}{1 - \alpha^{2}} \right) , \nonumber \\
S_1 &=& - 2 \alpha k^{2} \left(\frac{1 + \alpha^{2}}{1 - \alpha^{2}} \right)
= J = M a , \\
M_2 &=& - k^{3} \left(q_\text{{\sc mn}}
+ 4 \alpha^{2} \frac{(1 + \alpha^{2})}{(1 - \alpha^{2})^{3}} \right)
= - k^{3} q_\text{{\sc mn}} - M a^{2} , \nonumber
\end{eqnarray}
from which we see that $q_\text{{\sc mn}} := q/k^3$

This version of the metric incorporates the corrections done by \cite{PhysRevD.77.024035} and \cite{PhysRevD.81.124005}.

\subsection{Expanded Hartle--Thorne (HT) metric} \label{a: HT}
The metric as given in \cite{1968ApJ...153..807H} was not expanded to the correct orders of accuracy (terms up to $J^2$ and $q$). Here we use the expanded version from \cite{Frutos2019}:

\begin{eqnarray}
\label{chtnew}
V &=& 1 - 2 \frac{M}{r} + 2 \frac{q}{r^3} P_2 + 2 \frac{M q}{r^4} P_2  - \frac{2}{3} \frac{J^2}{r^4} (2 P_2 + 1) ,
\nonumber \\
W &=& - 2 \frac{J}{r} \sin^2{\theta} -\frac{Jq}{r^4}P_3^1\sin\theta  , \nonumber \\
X &=& \left(1 - 2 \frac{M}{r} \right)^{-1} \left[
1 - 2 \frac{q}{r^3} P_2 - 6 \frac{M q}{r^4} P_2 \right. \nonumber \\
&+& \left. 2 \frac{J^2}{r^4} (8 P_2 - 1) \right] , \\
Y &=& r^2 \left[1 - 2 \frac{q}{r^3} P_2 - 5 \frac{M q}{r^4} P_2
+ \frac{J^2}{r^4} P_2 \right] , \nonumber \\
Z &=& r^2 \sin^2{\theta} \left[1 - 2 \frac{q}{r^3} P_2 - 5 \frac{M q}{r^4} P_2
+ \frac{J^2}{r^4} P_2 \right]  \nonumber
\end{eqnarray}

where $P_3^1$ is the associated Legendre polynomial $P_3^1 = (5P_2 + 1)\sin\theta$. We discovered that the second term in $W$ can be omitted without any significative differences in the results for the configurations tested, while gaining a 13 per cent speed up in execution time.

\subsection{Metric in Frutos (2016) (Fru16)} \label{a: Fru16}

With the mass quadrupole moment $ q_\text{{\sc f}} := -q $, the metric potentials are given by

\begin{equation}
\begin{aligned}
V &= -\frac{e^{-2\psi}}{\rho^2} [a^2\sin^2\theta - \Delta]\\
W &= \frac{a}{\rho^2}[ \Delta - (r^2 + a^2) ]\sin^2\theta + \frac{Jq_\text{{\sc f}} }{r^4}P_3^1\sin\theta \\
X &= \rho^2 \frac{e^{2\chi}}{\Delta}\\
Y &= \rho^2 e^{2\chi}\\
Z &= \frac{e^{2\psi}}{\rho^2} [ (r^2+a^2)^2 - a^2\Delta \sin^2\theta ]\sin^2\theta
\end{aligned}
\end{equation}

where

\begin{align}
\Delta &= r^2 - 2Mr +a^2 \\
\rho^2 &= r^2 + a^2\cos^2 \theta \\
\psi &= \frac{q_\text{{\sc f}}}{r^3}P_2 + 3 \frac{Mq_\text{{\sc f}}}{r^4}P_2 \label{eq:metrics-frutos-psi}\\
\chi &= \frac{q_\text{{\sc f}}P_2}{r^3} + \frac{Mq_\text{{\sc f}}}{r^4}\left( -\frac{1}{3} + \frac{5}{3}P_2 + \frac{5}{3}P_2^2 \right) \notag\\
&\quad+ \frac{q_\text{{\sc f}}^2}{r^6}\left(\frac{2}{9}-\frac{2}{3}P_2 -\frac{7}{3}P_2^2 + \frac{25}{9}P_2^3 \right)
\end{align}

The first three relativistic multipoles are giving by

\begin{eqnarray}
\label{multipoles}
M_0 &=& M , \nonumber \\
S_1 &=& J = M a , \\
M_2 &=& q_\text{{\sc f}} - M a^{2} . \nonumber
\end{eqnarray}

in a similar way than the Hartle--Thorne metric, the second term in $W$ can be omitted, resulting in a small gain in performance without significant losses in accuracy.

\subsection{Glampedakis--Babak (GB) metric} \label{ap: GB}

This metric was given in \cite{2006CQGra..23.4167G} in the form $g_{\alpha \beta} = g_{\alpha\beta}^\text{Kerr} + \epsilon h_{\alpha\beta}$, with the second term being the corrections from the Abramowicz form of the Hartle--Thorne metric. However, in the paper, only the contravariant components of the metric were given. We inverted them up to the correct order, obtaining the potentials

\begin{eqnarray}
\label{GB}
V &=& \left(1 - 2 \frac{M r}{\rho^2} \right)
+ 2 \epsilon P_2 F_1 \left(1 - 2 \frac{M}{r} \right) , \nonumber \\
W &=& - 2 \frac{J r}{\rho^2} \sin^2{\theta} , \nonumber \\
X &=& \frac{\rho^2}{\Delta}
- 2 \epsilon P_2 F_1 \left(1 - 2 \frac{M}{r} \right)^{-1} , \\
Y &=& {\rho^2} + 2 \epsilon r^2 P_2 F_2 . \nonumber \\
Z &=& \left(r^2 + a^2 + 2 \frac{M a^2 r}{\rho^2} \sin^2{\theta} \right)
\sin^2{\theta} + 2 \epsilon r^2 P_2 F_2 \sin^2{\theta} . \nonumber
\end{eqnarray}

\noindent
with

\begin{eqnarray}
\label{funcs1}
\Delta &=& r^2 - 2 M r + a^2 , \nonumber \\
\rho^2 &=& r^2 + a^2 \cos^2{\theta} , \\
P_2 &=& \frac{1}{2} {(3 \cos^2{\theta} - 1)} . \nonumber
\end{eqnarray}

\noindent
The function $ P_2 $ is a Legendre polynomial. The parameter $ \epsilon $ is related with the quadrupole moment through $ \epsilon = - q/M^3 $. The functions $ F_1 $ and $ F_2 $ are given by

\begin{eqnarray}
\label{funcs2}
F_1 &=& - 5 (r - M) \frac{(2 M^2 + 6 M r - 3 r^2)}{(8 M r (r - 2 M))}\\
&+& \frac{15}{16} \frac{r(r - 2 M)}{M^2} \ln{\left(1 - 2 \frac{M}{r} \right)} ,
\nonumber \\
F_2 &=& \frac{5}{8} \frac{(2 M^2 - 3 M r - 3 r^2)}{M r} \\
&-& \frac{15}{16} \frac{(r^2 - 2 M^2)}{M^2} \nonumber
\ln{\left(1 - 2 \frac{M}{r} \right)} .
\end{eqnarray}

\bsp	
\label{lastpage}

\end{document}